\documentclass[12pt]{iopart} %
\usepackage{amssymb}
\expandafter\let\csname equation*\endcsname\relax

\expandafter\let\csname endequation*\endcsname\relax
\usepackage{amsmath}
\usepackage{mathrsfs}
\usepackage{graphicx}
\usepackage{upgreek}
\usepackage{hyperref}
\usepackage{xcolor}
\usepackage{orcidlink}
\usepackage{float}

 \newcommand{\BEQ}{\begin{equation}}     %
\newcommand{\BEA}{\begin{eqnarray}}
\newcommand{\BD}{\begin{displaymath}}
\newcommand{\EEQ}{\end{equation}}       %
\newcommand{\EEA}{\end{eqnarray}}
\newcommand{\ED}{\end{displaymath}}
\newcommand{\lap}[1]{\overline{#1}}

\newcommand{\vep}{\varepsilon}          %
\newcommand{\D}{{\rm d}}                %
\newcommand{\II}{{\rm i}}      
\newcommand{\erf}{{\rm erf\,}}          %
\newcommand{\wht}[1]{\widehat{#1}}      %
\renewcommand{\lap}[1]{\overline{#1}}     %

\newcommand{\appsection}[2]{\setcounter{equation}{0}\setcounter{subsection}{0}\setcounter{figure}{0}
\section*{Appendix #1. #2}
\renewcommand{\theequation}{#1.\arabic{equation}}\renewcommand{\thefigure}{#1.\arabic{figure}}
              \renewcommand{\thesection}{#1} }

\makeatletter
\newcommand*{\centerfloat}{%
  \parindent \z@
  \leftskip \z@ \@plus 1fil \@minus \textwidth
  \rightskip\leftskip
  \parfillskip \z@skip}
\makeatother

\begin{document}

\title{Fractional diffusion equations interpolate between damping and waves}

\author{Andy Manapany$^{a,b}$, S\'ebastien Fumeron$^a$, Malte Henkel$^{a,c}$}

\address{$^a$ Laboratoire de Physique et Chimie Th\'eoriques (CNRS UMR 7019),
 Universit\'e de Lorraine Nancy, B.P. 70239, F -- 54506 Vand{\oe}uvre-l\`es-Nancy Cedex, France\\
 $^b$ Applied Mathematics Research Centre, Coventry
University, Coventry CV1 5FB, United Kingdom\\
 $^c$ Centro de F\'{i}sica Te\'{o}rica e Computacional, Universidade de Lisboa, \\Campo Grande, P--1749-016 Lisboa, Portugal\\

}
\ead{andy.manapany@univ-lorraine.fr, sebastien.fumeron@univ-lorraine.fr, malte.henkel@univ-lorraine.fr}
\vspace{10pt}
\begin{indented}
\item[] \today
\end{indented}

\begin{abstract}
The behaviour of the solutions of the time-fractional diffusion equation, based on the Caputo derivative, is studied and its dependence on the fractional exponent is analysed. 
The time-fractional convection-diffusion equation is also solved and an application to Pennes bioheat model is presented. Generically, a wave-like transport at short times passes over to a 
diffusion-like behaviour at later times. 
\end{abstract}

\vfill

\setcounter{footnote}{0}

\section{Introduction}

Fractional calculus has now become a widespread tool used in various physical contexts, characterised by strong effects of non-locality and/or memory, with examples
ranging from biophysics \cite{magin2004fractional,magin2010fractional,ezzat2014fractional,valentim2020can,monteiro2021fractional} to material science 
\cite{koeller_applications_1984,alcoutlabi1998application,meral2010fractional,stoimenov2013non,failla2020advanced} 
and engineering \cite{loverro2004fractional,gutierrez2010fractional,tenreiro2010some,sun2018new}. 
Diffusion makes no exception: while in its classical form it results from brownian random walks with short-ranged and statistically uncorrelated steps, 
fractional diffusion has emerged as a powerful tool for 
the description of processes involving anomalous diffusion, long-range interactions and memory effects. 
Indeed, fractional calculus was applied to a wide range of diffusion problems including non-Fickian models \cite{ramirez-torres_influence_2021,xu2022non,matias2023model,Henkel2016}, 
Fokker-Planck equations \cite{metzler1999deriving,yanovsky2000levy, barkai2001fractional, liu2004numerical,li2020anomalous} 
and master equations \cite{hilfer1995fractional,Henkel02,tarasov2009fractional,pagnini2012generalized}.

{}From a mathematical standpoint, fractional calculus goes back at least two centuries \cite{h_solution_1823}. 
Its cornerstone is the Riemann-Liouville integral ({\sc rli}), which generalises the well-known 
Cauchy formula for $n$-fold (repeated) integrations to the case of non-integer values of $n$. If $\beta$ is a real number, the {\sc rli} is defined as 
\begin{equation} 
    I^{\beta}_a f(x) = \frac{1}{\Gamma(\beta)}\int_{a}^{x} \!\D \tau\: (x-\tau)^{\beta-1} f(\tau) \label{RLI}
\end{equation}
where $\Gamma$ is Euler's Gamma function. For $\beta=n\in\mathbb{N}$, this reduces to Cauchy formula. 
To obtain a fractional derivative, one might wish to consider negative values of $\beta$, but care is needed in order to avoid divergent integrals. 
Commonly, a fractional derivative of order $\alpha=1-\beta$ of a function $f(x)$ can  be defined in the Riemann-Liouville fashion as $\frac{\D^n}{\D x^n}\bigl( I^{1-\alpha}_a f(x)\bigr)$. Alternatively, which is the choice we shall adopt throughout this paper, one may use the {\em Caputo fractional derivative} 
\begin{equation} \label{eq:Caputo}
    \partial_x^{\alpha} f(x) = D^\alpha_{0,x}f(x) :=\frac{1}{\Gamma(n-a)}\int_{0}^{x} \!\D\tau\: (x-\tau)^{n-\alpha-1}f^{(n)}(\tau) 
\end{equation}
where $\alpha$ is the fractional order of diffusion, and $n=[\alpha]+1$ is the next integer above it. In addition, $f^{(n)}(x)=\frac{\D^n f(x)}{\D x^n}$. 
The Caputo fractional derivative (\ref{eq:Caputo})  has the property $\partial_x^{\alpha} 1 =0$
which distinguishes it from many other alternative definitions.\footnote{The literature abounds with different and in-equivalent definitions, each one having their own assets and drawbacks.}

Applications of fractional calculus begin with addressing the following issues:
\begin{enumerate}
    \item which definition of fractional derivatives is the best suited to a given problem~? 
    \item how can fractional derivatives properly implemented in a given model~? 
\end{enumerate}
Clearly, there cannot be an universally applicable {\it a priori} answer to these questions. 
Traditionally, the Riemann-Liouville definition has been widely used, but more recently, physicists are coming to prefer
the Caputo definition (\ref{eq:Caputo}) since it is particularly well-suited for the statement of  Cauchy initial-value problems \cite{kilbas2004cauchy}. 
To address the second point, several strategies have been used, notably on when to switch from integer-order derivative to its fractional counterpart 
(for a discussion, see section~3 in \cite{fumeron2023fractional}). 

While there is a rich literature on ordinary fractional differential equations \cite{gorenflo1997fractional, podlubny_fractional_1998,kilbas_theory_2006}, 
considerably less is known for fractional partial differential equations \cite{gorenflo2000wright,mainardi2001fundamental}. We shall explore the behaviour of the solutions of
a simple $1D$ fractional diffusion equation, with the Caputo derivative in time. The availability of an exact solution, in terms of Mittag-Leffler functions, 
simplifies our study of the dependence of the solution's behaviour on the
fractional order $\alpha$, notably how it interpolates between a classical diffusive behaviour ($\alpha=1$) 
and a propagating wave ($\alpha=2$). A time-dependent cross-over between two distinct regimes is found.  
The last section is dedicated to an application to biophysics and proposes a simplified model for the thermal ablation of tumorous cells. 
Two appendices contain material on the Mittag-Leffler function and review several
simple techniques for the solution of partial differential equations. 

\section{Simple 1D fractional diffusion equation} \label{sec2} 

Partial differential equation are often treated by reducing them to ordinary differential equations. 
Here we begin by adapting the time-honoured technique of separation of variables in linear partial differential equations to their fractional
counter-part. Well-established techniques \cite{Mainardi1996,podlubny_fractional_1998} 
are used to solve the associated ordinary equations (see appendix~B for alternative techniques for finding solutions in semi-infinite space). 
Throughout, we shall frequently need properties of the Mittag-Leffler function $E_{\alpha,\beta}(z)$, which are collected in appendix~A.

We concentrate on the simple $1D$ fractional diffusion equation, for a function $T=T(t,x)$ thought to represent a time- and space-dependent temperature 
\begin{equation}\label{diff}
    \partial_t^{\alpha} T =D\frac{\partial^2 T}{\partial x^2}
\end{equation}
with the diffusion constant $D>0$ and the Caputo derivative (\ref{eq:Caputo}). For $\alpha=1$, this reduces to the usual diffusion equation, 
while for $\alpha=2$ this becomes a wave equation where $D=\frac{1}{c^2}$ is related to the wave propagation velocity $c$. 
To be specific, we consider the following initial and boundary conditions
\begin{equation}  \label{inibord}
    \begin{cases}
    T(t,0)=0\\
    T(t,L)=0 \\
    T(0,x)=\frac{100x}{L}
    \end{cases}
\end{equation}
which confine our system to the spatial range $0\leq x \leq L$. 
In spite of its innocent-looking appearance, the fractional derivative implies that (\ref{diff}) really is a non-local partial
integro-differential equation. Still, because (\ref{diff}) is linear, it turns out that the explicit solution can be constructed via a formal analogy with the
theory of partial differential equations. 
One therefore may try a separation of variables $T(t,x) = f(t) g(x)$ such that
(\ref{diff}) decouples into two equations, each involving only one of the independent variables, namely 
\begin{subequations}
\begin{align}
\label{time}
    \frac{\partial_t^{\alpha} f(t)}{f(t)} &=-k^2 \\ 
\label{space}
    \frac{g''(x)}{g(x)} &=-k^2
\end{align}
\end{subequations}
where $k$ is the separation constant and $g'(x)=\frac{\D g(x)}{\D x}$ is the ordinary derivative. 

The spatial part (\ref{space}) readily gives $g(x) = g_0 \sin(kx) + g_1 \cos(kx)$. 
The temporal part (\ref{time}) is conveniently solved via a Laplace transform 
$\lap{f}(p) = \mathscr{L}\bigl( f(t)\bigr)(p) = \int_0^{\infty} \!\D t\: e^{-pt} f(t)$. 
For the Caputo derivative, it is known that \cite{Podlubny1997,Haubold11,Gorenflo14}
\begin{equation} \label{eq:LapCap}
    \int_{0}^{\infty} \!\D t\: e^{-pt} \partial_t^{\alpha}f(t) = p^\alpha \lap{f}(p)-\sum_{k=0}^{[\alpha]} p^{\alpha-k-1}f^{(k)}(0)
\end{equation}
It is specific to the Caputo derivative (\ref{eq:Caputo}) that its initial data are the ordinary derivatives at the origin. 
That is not so for different fractional derivatives (e.g. the Riemann-Liouville fractional derivative). 

First, we restrict attention to the case $0<\alpha<1$. Then $n=[\alpha]+1=1$ and (\ref{time}) turns into 
\begin{equation}
    p^\alpha \lap{f}(p)-p^{\alpha-1}f(0)+Dk^2 \lap{f}(p)=0
\end{equation}
or equivalently 
\begin{equation}
    \lap{f}(p)=\frac{p^{\alpha-1}f(0)}{p^\alpha+D k^2}
\end{equation}
Using the identity (\ref{eq:lapML}) this can be inverted via the Mittag-Leffler function and one has 
\begin{equation}\label{f}
    f(t)=f(0) E_{\alpha,1}(-D k^2 t^\alpha)
\end{equation}
One now combines with the spatial solution $g(x)$ given above and has the formal solution 
\begin{equation} \label{eq:ansatzT1}
    T(t,x)= \sum_{k} \left[ f_0(k) E_{\alpha,1}(-D k^2 t^\alpha)\sin(kx) + f_1(k) E_{\alpha,1}(-D k^2 t^\alpha)\cos(kx)\right] 
\end{equation}
where the sum extends over all `admissible' values of $k$. This reflects general theorems on the number of solutions of fractional differential equations \cite{Diethelm10}. 
The boundary/initial conditions (\ref{inibord}) will now be used to specify these and also fix the constants $f_{0,1}(k)$. 

The first two boundary conditions (\ref{inibord}) hold true for all times. Inserting the first one into the formal
solution (\ref{eq:ansatzT1}), it follows that $f_1(k)=0$. Then, the second one is satisfied for all values of $k$ such that
$\sin kL =0$, which means that $k=\pi m$ with $m\in\mathbb{N}$. Then one has
\begin{equation} \label{eq:ansatzT2}
T(t,x) = \sum_{m=1}^{\infty} f_m E_{\alpha,1}\left( -D \left(\frac{\pi m}{L}\right)^2 t^{\alpha} \right) \sin\left( \pi m \frac{x}{L}\right)
\end{equation}
and the remaining constants $f_m$ are now found from the last initial condition (\ref{inibord}).  At the initial time $t=0$, 
(\ref{eq:ansatzT2}) becomes a Fourier series for the function $\frac{100 x}{L}$
\begin{equation}
    T(0,x)=\sum_{m=1}^{\infty} f_m  \sin\left(\frac{m\pi}{L}x\right) = \frac{100x}{L}
\end{equation}
in the period interval $[0,L]$. The Fourier coefficients $f_n$ are found as usual 
\begin{equation}
    \begin{split}
        f_n & =\frac{2}{L}\int_{0}^{L} \frac{100}{L} x \sin\left(\frac{n\pi}{L}x\right) \D x\\
        & = \frac{200}{L^2} \bigg\{ \left[-\frac{L}{n\pi}x\cos\left(\frac{n\pi}{L} x\right)\right] _{0}^{L} + \frac{L}{n\pi}\int_{0}^{L}   \cos\left(\frac{n\pi}{L}x\right) \D x \bigg\}\\
        & = \frac{200}{n \pi} (-1)^{n-1}
    \end{split}
\end{equation}
which gives the final solution to the system (\ref{diff},\ref{inibord}), for any $0<\alpha\leq 1$ 
\begin{equation} \label{eq:Tdiff-sol}
    T(t,x)=\frac{200}{\pi}\sum_{n=1}^{\infty} \frac{(-1)^{n-1}}{n} E_{\alpha,1}\left(-D \frac{n^2\pi^2}{L^2} t^\alpha\right)\sin\left(\frac{n\pi}{L}x\right)
\end{equation}

Next, we look for the extension of this result in the range $1<\alpha\leq 2$. Then (\ref{eq:LapCap}) takes the form
\begin{equation}
    \int_{0}^{\infty} \!\D t\: e^{-pt} \partial_t^{\alpha}f(t) =p^\alpha \lap{f}(p) - p^{\alpha-1}f(0)-p^{\alpha-2}f^{'}(0)
\end{equation}
The separation of variables goes through essentially as before, but now it follows from (\ref{time}) that 
\begin{equation}
    \lap{f}(p)=\frac{p^{\alpha-1}f(0)}{p^\alpha+D k^2}+\frac{p^{\alpha-2}f'(0)}{p^\alpha+D k^2}
\end{equation}
Using the linearity of the Laplace transform, this is readily inverted using (\ref{eq:lapML}) to give 
\begin{equation}
    f(t)=f(0)E_{\alpha,1}(-D k^2 t^\alpha) + f'(0)tE_{\alpha,2}(-D k^2 t^\alpha)
\end{equation}
and with respect to eq.~(\ref{f}) contains a further term. 
Indeed, two independent solutions are generically expected for $1<\alpha\leq 2$ \cite{Diethelm10}. 
Then the formal solution of (\ref{diff}) becomes
\begin{eqnarray} 
    T(t,x) &=& \sum_{k} \left[ f_0(k) E_{\alpha,1}(-D k^2 t^\alpha)\sin(kx) + f_1(k) E_{\alpha,1}(-D k^2 t^\alpha)\cos(kx)\right. \nonumber \\
    & & + \left. f_0'(k) t E_{\alpha,2}(-D k^2 t^\alpha)\sin(kx) + f_1'(k) t E_{\alpha,2}(-D k^2 t^\alpha)\cos(kx)\right]  \label{eq:ansatzT3} \nonumber \\
    & & 
\end{eqnarray}
where the constants $f_{0,1}(k), f_{0,1}'(k)$ must be found from the initial and boundary conditions 
(\ref{inibord}). Form the first one, and since $E_{\alpha,1}(t^{\alpha})$ and $t E_{\alpha,2}(t^{\alpha})$
are linearly independent because of the recurrences (\ref{ML-recur}), it follows that $f_1(k)=f_1'(k)=0$. 
{}From the second one, one gets the requirement that $\sin kL=0$, hence $kL = m \pi$ with $m\in\mathbb{N}$. 
This leads to 
\BEQ 
    T(t,x) = \sum_{m=1}^{\infty} \left[ f_m E_{\alpha,1}\left(-D \left(\frac{\pi m}{L}\right)^2 t^\alpha\right) 
    + f_m' t E_{\alpha,2}\left(-D \left(\frac{\pi m}{L}\right)^2 t^\alpha\right) \right]  \sin\left(\frac{\pi m}{L}x\right)~~~ \label{eq:ansatzT4}
\EEQ
The last initial condition (\ref{inibord}) only fixes $f_m = \frac{200}{\pi} \frac{(-1)^{m-1}}{m}$ 
whereas $f_m'$ remains undetermined. For its specification, a further initial condition, specifying e.g. the velocity of
the initial distribution, must be given. If for example we assume that initially, the starting configuration
is stationary, that is $\partial_t T(0,x)=0$  we find
\begin{equation}
\left. \partial_t T(t,x) \right|_{t=0} = 0 = \sum_{m=1}^{\infty} f_m'  \sin\left(\frac{\pi m}{L}\right)
\end{equation}
from which it follows that $f_m'=0$.

We conclude that for an initial configuration assumed stationary, {\em the validity of the 
solution (\ref{eq:Tdiff-sol}) is extended to the whole range $0<\alpha\leq 2$.} 
Of course, this is but an example illustrating more general theorems \cite{Diethelm10}.

We can also illustrate this method on another example, using the following initial conditions instead:
\begin{equation} \label{21}
    \begin{cases}
    T(0,x)=T_0(x)=\sin{(\frac{\pi x}{L})}\\
    \frac{\partial T}{\partial t}(0,x)=\frac{\pi}{L}\cos{\frac{\pi x}{L}}\\
     T(t,0)=0\\
    T(t,L)=0
    \end{cases}
\end{equation}
Just as illustrated previously, taking into account the boundary conditions allows us to search for a solution. Carrying out once the separation of variables, we find the general solution 
\begin{eqnarray}\nonumber
    \hspace{-1.5cm} T(t,x) &=& \sum_{m=1}^{\infty} \Big[ \left(f_m E_{\alpha,1}\left(-D \left(\frac{\pi m}{L}\right)^2 t^\alpha\right) 
    + f_m' t E_{\alpha,2}\left(-D \left(\frac{\pi m}{L}\right)^2 t^\alpha\right) \right)  \sin\left(\frac{\pi m}{L}x\right)\\
    \hspace{-1.5cm}& & + \left(g_m E_{\alpha,1}\left(-D \left(\frac{\pi m}{L}\right)^2 t^\alpha\right) 
    + g_m' t E_{\alpha,2}\left(-D \left(\frac{\pi m}{L}\right)^2 t^\alpha\right) \right)  \cos\left(\frac{\pi m}{L}x\right) \Bigr]\nonumber
\end{eqnarray}
Using the same method, we decompose into Fourier sums both the initial conditions in order to derive the expressions of $f_m, f_m', g_m$ and $g_m'$. We then find :
\begin{equation}
    \begin{cases}
    \forall m \ge 2,  g_m = \frac{1-(-1)^{m+1}}{\pi (m+1)}+\frac{(-1)^{1-m}-1}{\pi (1-m)} \\
    g_0=\frac{2}{\pi}\\
    f_m = 0\\
    f_m'=\frac{2((-1)^{m+1}-1)}{L (m+1)}+\frac{2((-1)^{1-m}-1)}{L (1-m)}
    \end{cases}
\end{equation}
and we arrive at the explicit solution for the initially non-stationary profile (\ref{21}) 
\begin{eqnarray}\nonumber
    T(t,x) &=&\frac{2}{\pi}+ \sum_{n=2}^{\infty} \Bigl[\frac{1-(-1)^{m+1}}{\pi(m+1)}+\frac{(-1)^{m+1}-1}{\pi(1-m)} \Bigr]E_{\alpha,1}\left(-D \frac{m^2 \pi^2}{L^2} t^\alpha\right)\cos\left(\frac{m\pi}{L}x\right)\\ \nonumber
    & &+\Bigl[\frac{2(-1)^{m+1}-1}{L(m+1)}+\frac{2(-1)^{m+1}-1}{L(1-m)}\Bigr]t E_{\alpha,2}\left(-D \frac{m^2 \pi^2}{L^2} t^\alpha\right) \sin\left(\frac{m\pi}{L}x\right)\\\nonumber
\end{eqnarray}

\newpage
\section{Result of the fractional diffusion equation}\label{sec3}

We now study the behaviour of the $\alpha$-dependent solutions, derived in the last section, 
by plotting their spatial dependence for several values of the time $t$. 
However, it is not advisable to simply truncate the Fourier series at some finite order, since the partial sums 
$T_i=T_i(t,x)$ do not always uniformly converge for $i\to\infty$ \cite{courantjohn1965}. 
Notably, if the limit function $T_{\infty}(t,x)$ turns out to be discontinuous, Gibb's phenomenon
arises which leads to an overshoot of the approximate Fourier sum $T_i$ by about $9\%$ of the jump height
of $T_{\infty}(t,x)$, at the discontinuity. 
This can be cured by using the Fejer representation \cite{courantjohn1965}  
\begin{equation}
    F_n(t,x)=\sum_{i=0}^{n} \frac{T_i(t,x)}{n+1}
\end{equation}
where $T_i(t,x)$ stands for the Fourier sum up to, and including, the $i^{\rm th}$ term. The Fejer sums
$F_n$ converge uniformly as $n\to\infty$. 

\vspace{10mm}
\begin{figure}[tb]
    \centerfloat
    \includegraphics[scale=0.25]{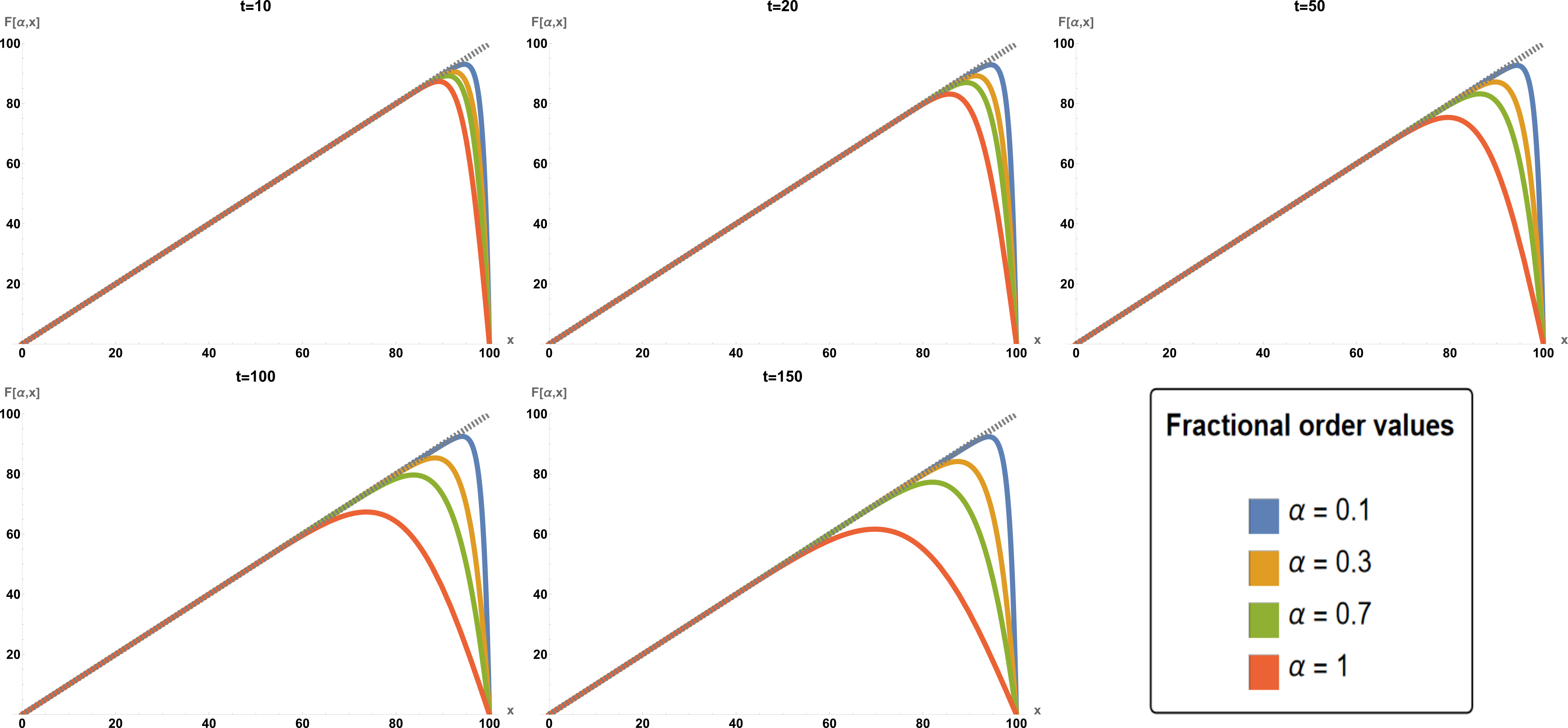}
    \caption[fig 1]{Temperature profile $T(t,x)$ at times $t=[10,20,50,100,150]$ for low values of the fractional order 
    $\alpha=[0.1,0.3,0.7,1]$ from top to bottom. The initial profile is also indicated.}
    \label{fig:shift1}
\end{figure}

\begin{figure}[tb]
    \centerfloat
    \includegraphics[scale=0.25]{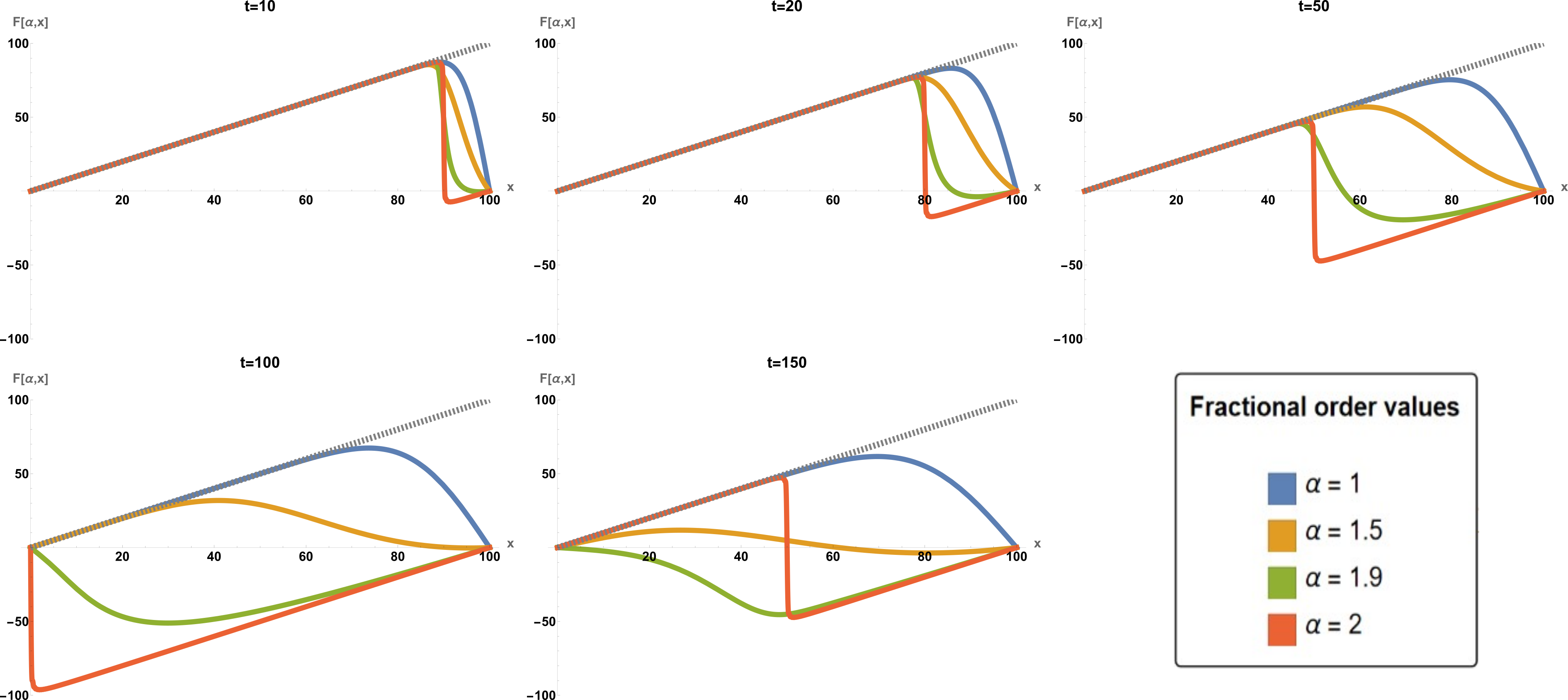}
    \caption[fig2]{Temperature profile $T(t,x)$ at simulation times $t=[10,20,50,100,150]$ f
    or higher values of the fractional order $\alpha=[1,1.5,1.9,2]$ from top to bottom at the right of the panels. 
    The initial profile is also indicated. 
    }
    \label{fig:shift2}
\end{figure}

Our first example concerns the evolution of a (periodically continued) linear profile with a discontinuity at the right boundary. The exact solution, for $0<\alpha\leq 2$ and stationary initial conditions, is given  by the Fourier series (\ref{eq:Tdiff-sol}). 
Figure~\ref{fig:shift1} shows the evolution for small values 
of the fractional parameter, in the interval $0<\alpha\leq 1$, whereas figure~\ref{fig:shift2} displays the evolution in the range $1\leq \alpha\leq 2$. 
The spatial profiles for the range $0<\alpha\leq 1$ all look qualitatively similar and all undergo an analogous evolution in the sense that the
initial profile is smoothed out with increasing time. However, when $\alpha$ is decreased from the
value $\alpha=1$ of simple diffusion, the fractional evolution becomes ever more slowly as $\alpha$ is decreased, see figure~\ref{fig:shift1}. 
On the other hand, in figure~\ref{fig:shift2} one sees the interpolation between simple diffusion when $\alpha=1$ and wave motion when $\alpha=2$. 
Increasing $\alpha$ beyond the value of simple diffusion, not only the evolution is becoming ever more fast, but there is as well an increasing 
tendency that the initial profile should be rigidly transported to the left and an associated damping is getting weaker as $2-\alpha\to 0^+$. 
Should one interpret this as a suggestion that a cross-over between different types of dynamical behaviour, namely diffusive and wavelike, occurs~? 
In the last panel in figure~\ref{fig:shift2}, at time $t=150$, one sees that for the wave-motion with $\alpha=2$, the signal was reflected
at the left boundary and returns, whereas for the other values with $\alpha<2$, the left boundary has not yet been reached.

\begin{figure}[tb]
    \centerfloat
    \includegraphics[scale=0.4]{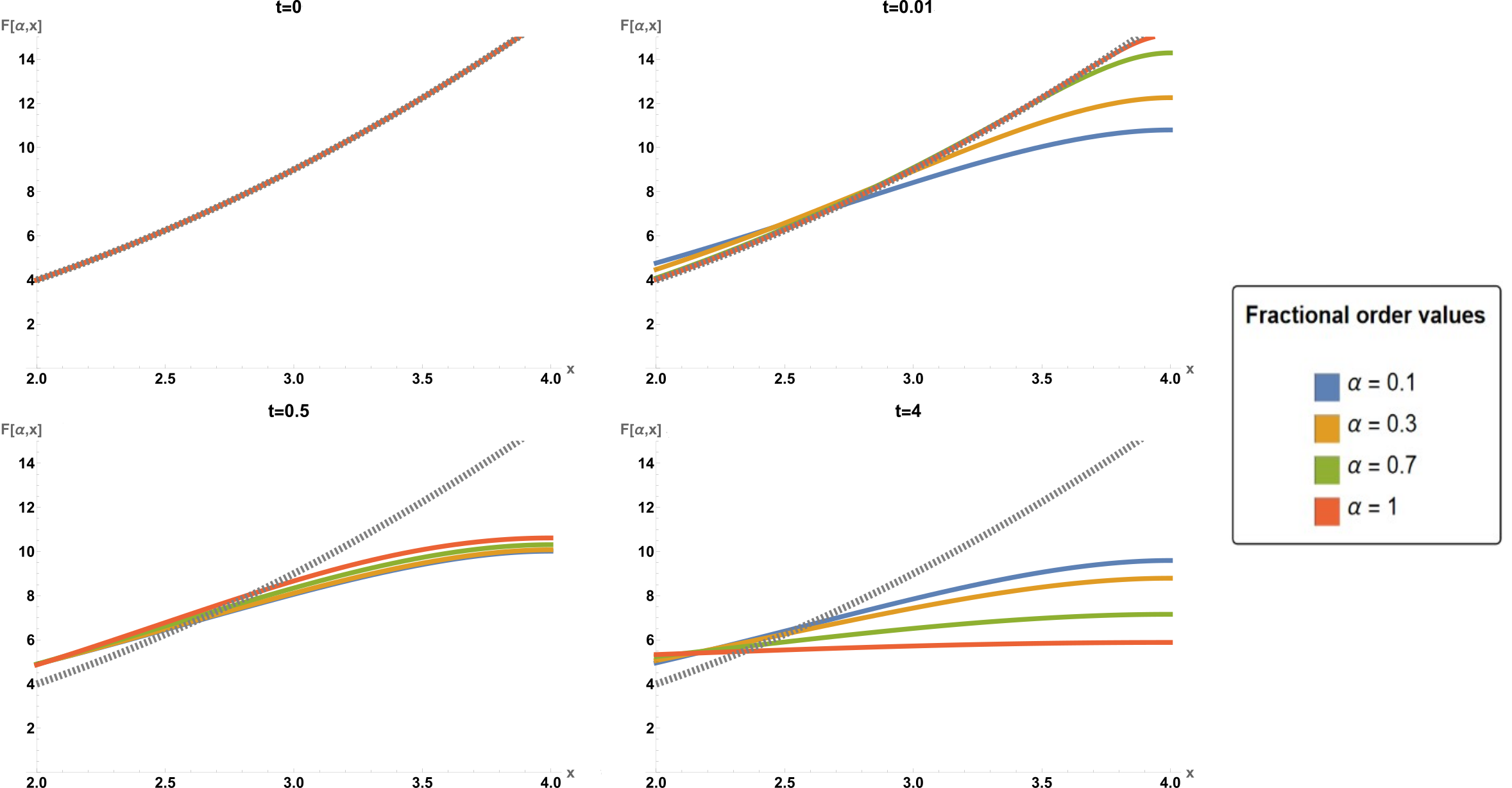}
    \caption[fig3]{Temperature profile $T(t,x)$ at times $t=[0,0.01,0.5,4]$ 
    for a quadratic initial profile and low values of the fractional order $\alpha=[0.1,0.3,0.7,1]$. 
    The initial profile is also indicated.
    }
    \label{fig:shift3}
\end{figure}

\begin{figure}[tb]
    \centerfloat
    \includegraphics[scale=0.4]{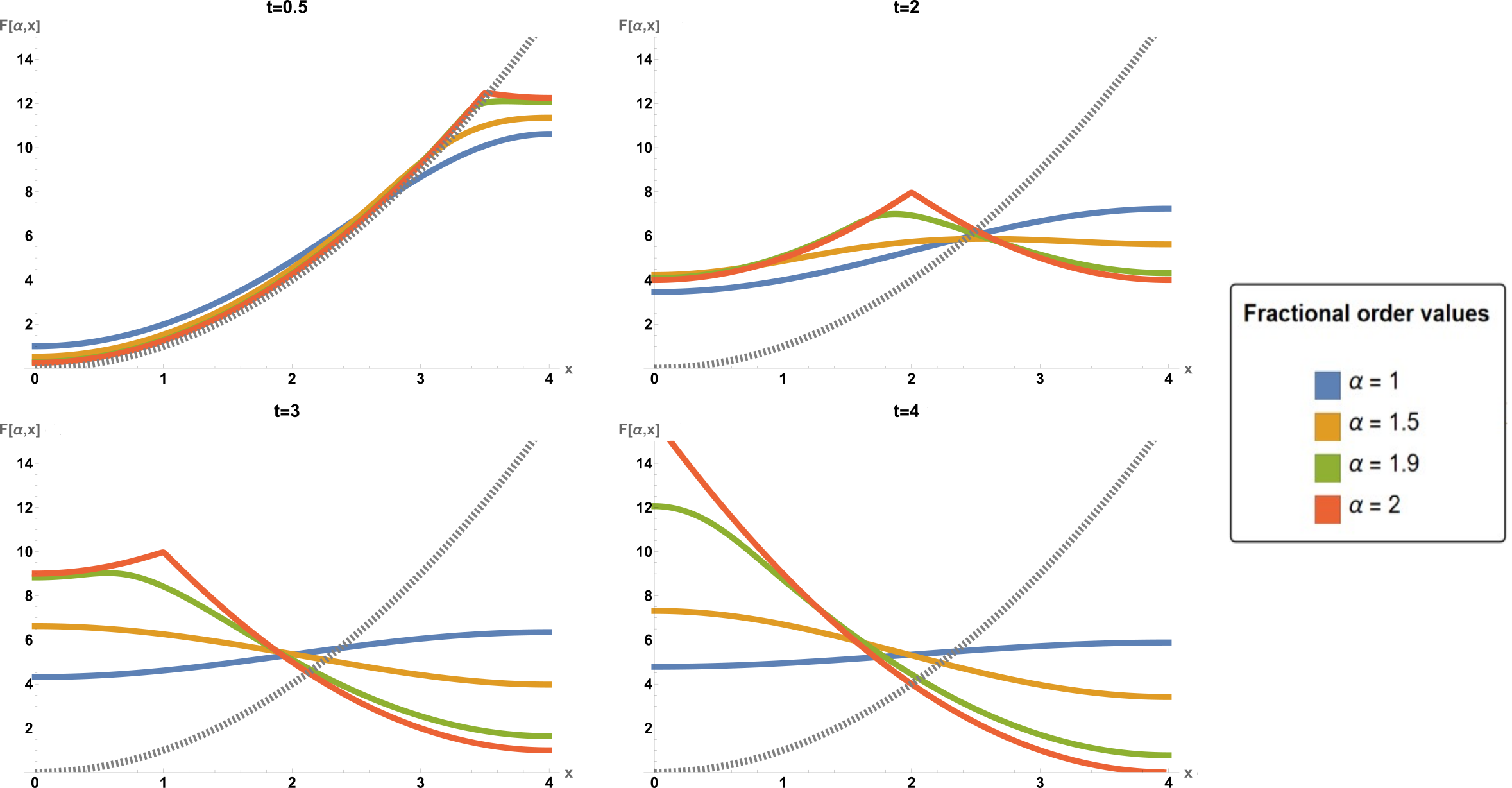}
    \caption[fig4]{Temperature profile $T(t,x)$ at times $t=[0.5,2,3,4]$ for a quadratic initial profile and higher values of the fractional order 
    $\alpha=[1,1.5,1.9,2]$. The initial profile is also indicated.
    }
    \label{fig:shift4}
\end{figure}

In order to present these effects more clearly, we now consider two further initial profiles. 
Our second example is a quadratic initial profile, such that the exact solution
has the Fourier series representation
\begin{equation} \label{x2}
    T(t,x) =\frac{L^2}{3}+ \sum_{n=2}^{\infty} \left[\frac{4 L^2}{(n \pi)^2} (-1)^{n}E_{\alpha,1}\left(-D \frac{n^2 \pi^2}{L^2} t^\alpha\right)
    \right] \cos\left(\frac{n\pi}{L}x\right)
\end{equation}
In figures~\ref{fig:shift3} and~\ref{fig:shift4}, we first display the evolution of a quadratic initial profile. In the regime $0<\alpha\leq 1$ 
of diffusive-like motion, we observe first from figure~\ref{fig:shift3} that for very small
times $t$, there is a very rapid evolution for the {\em smallest} values of $\alpha$. For slightly
larger times, all profiles almost re-collapse onto the same curve at some time $t=t_{\times}$, 
which is quite distinct from the initial profile (panel for $t=0.5$ in figure~\ref{fig:shift3}. 
Only for times $t>t_{\times}$ we recover an usual diffusive-looking motion which becomes faster when $\alpha$ increases towards $1$. 
\begin{figure}[tb]
    \centerfloat
    \includegraphics[scale=0.4]{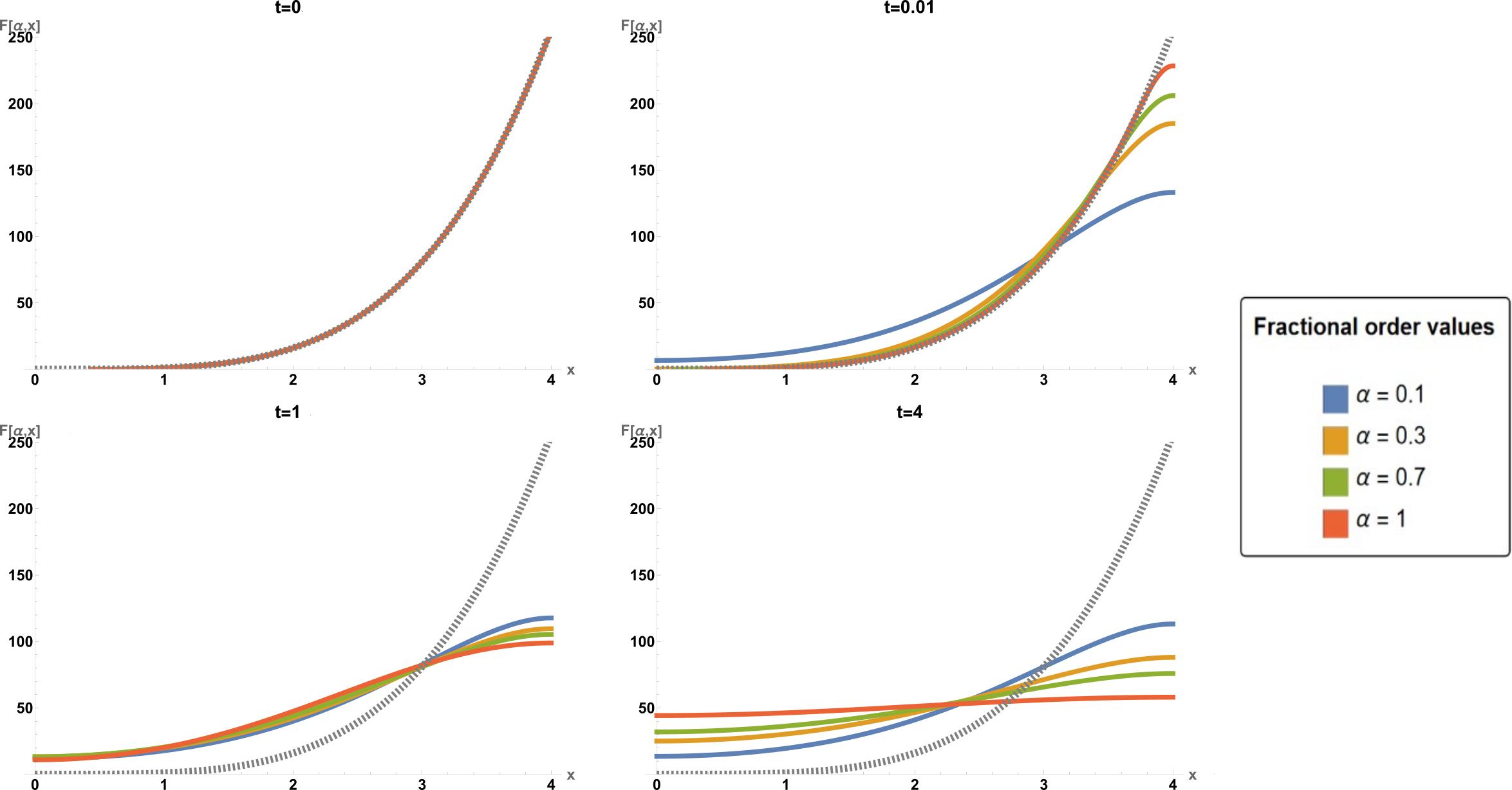}
    \caption[fig5]{Temperature profile $T(t,x)$ at time $t=[0,0.01,1,4]$ 
    for a quartic initial profile and low values of the fractional order $\alpha=[0.1,0.3,0.7,1]$. The initial
    profile is also indicated. 
    }
    \label{fig:shift5}
\end{figure}
For the regime $1\leq\alpha\leq 2$ which interpolates between simple diffusion and wave-motion, see figure~\ref{fig:shift4}, we again find that for very short times, the motion is more rapid
for smaller values of $\alpha$. On the other hand, the initial profile is rigidly transported to the left when $\alpha$ is close enough to $2$. 
For larger values of $t$, one sees that the initial profile 
(almost) re-builds at the left boundary. Although there is quite some dissipation, even for a
value as small as $\alpha=1.5$, we still observe at least a tendency for re-building. For larger
times the profiles will be reflected at the left boundary and then shall move towards the right.

\begin{figure}[tb]
    \centerfloat
    \includegraphics[scale=0.4]{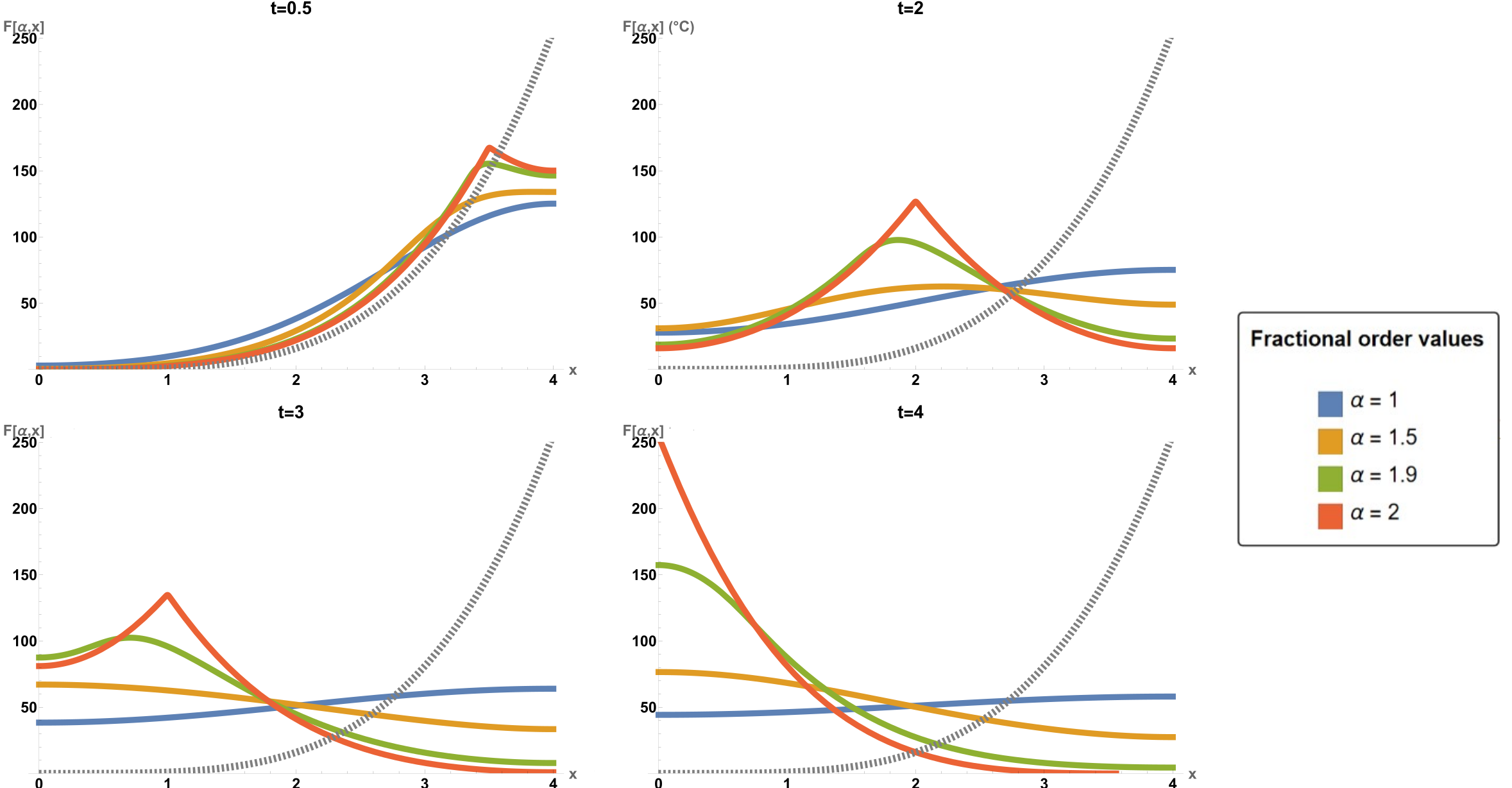}
    \caption[fig6]{Temperature profile $T(t,x)$ at times $t=[0.5,2,3,4]$ 
    for a quartic initial profile and higher values of the fractional order $\alpha=[1,1.5,1.9,2]$. The initial
    profile is also indicated. 
    }
    \label{fig:shift6}
\end{figure}

Our last example uses a quartic initial profile which leads to the following exact solution
\begin{equation}\label{x4}
    T(t,x) =\frac{L^4}{5}+ \sum_{n=2}^{\infty} \left[\Bigl(\frac{4 L^2}{(n \pi)^2}(-1)^{n}-\frac{48 L^4}{(n \pi)^4}(-1)^{n}\Bigl)E_{\alpha,1}\left(-D \frac{n^2 \pi^2}{L^2} t^\alpha\right)
    \right] \cos\left(\frac{n\pi}{L}x\right)   
\end{equation}
The corresponding fractional mouvements are shown in figures~\ref{fig:shift5} and~\ref{fig:shift6}, 
and further underscore the tendencies observed in the first two examples. In the region 
$0<\alpha\leq 1$, we observe from figure~\ref{fig:shift5} first that for very short times a rapid evolution for the smallest
values of $\alpha$, before at some later time $t=t_{\times}\approx 1$ there is an approximate re-collapse of all profiles, 
onto a curve distinct from the initial profile. Only for times $t>t_{\times}$ there is a qualitatively similar and diffusive-like mouvement. 
Things happen differently in the region $1\leq\alpha\leq 2$, see figure~\ref{fig:shift6}. 
Certainly, for very short times the evolution is the most fast for the most small values of $\alpha$. Then, for larger
times, the initial profile is rigidly transported to the left, albeit with some dissipation 
which becomes weaker as $2-\alpha\to 0^+$. For moderately large times, the initial profile builds
up again at the left boundary and even for the rather small value $\alpha=1.5$, at least the tendency for such a build-up is still visible.

Given the formal representations (\ref{eq:Tdiff-sol},\ref{x2},\ref{x4}) of the solutions in terms of Fourier series, 
it appears that they are distinguished by the increasing rapid decay of the Fourier
coefficient with the number $n$ of the modes. Going from the pair of figures~\ref{fig:shift1}, \ref{fig:shift2} 
of the first initial profile, on to the pair of figures~\ref{fig:shift3}, \ref{fig:shift4} and finally to the pair of 
figures~\ref{fig:shift5}, \ref{fig:shift6}, the number of modes which sensibly contribute to the entire sum becomes smaller. 
The time-dependence of each coefficient is given in terms of the Mittag-Leffler function $E_{\alpha,1}$.  

Eqs,~(\ref{eq:Tdiff-sol},\ref{x2},\ref{x4}) show that one needs the behaviour on the negative real axis of the Mittag-Leffler function 
$E_{\alpha,1}$ and we also refer to the figures~\ref{fig:MittagLeffler_agrand} and~\ref{fig:MittagLeffler_apetit} in appendix~A. From the asymptotic behaviour (\ref{eq:A:Easy}), we read off a generic algebraic decay for very large times. 
In addition, the monotonicity properties of $E_{\alpha,1}(-x)$
depend on whether $0<\alpha\leq 1$ or if $1<\alpha\leq 2$. In the first case, the Mittag-Leffler function is
always positive and decays monotonously to zero, see also figure~\ref{fig:MittagLeffler_apetit}, in agreement with the qualitatively diffusion-like behaviour
observed in figures~\ref{fig:shift1}, \ref{fig:shift3} and~\ref{fig:shift5}. 
On the other had, for $1<\alpha\leq 2$, one has for not too large arguments a regime with a (damped) oscillatory behaviour, see
(\ref{eq:A3c}) and figure~\ref{fig:MittagLeffler_agrand}. 
In the examples of quadratic or quartic initial profiles, since the Fourier coefficients 
fall off more rapidly, one the lowest modes will contribute such that the oscillatory regime will not be dominated effectively 
diffusion-like behaviour of the higher modes. Finally, since for very small arguments
the is a rapid fall-off of $E_{\alpha,1}(-x)$ which becomes stronger for smaller values of $\alpha$, 
the observations in figures~\ref{fig:shift3}, \ref{fig:shift4}, \ref{fig:shift5} and~\ref{fig:shift6} are explained as well. 

Formally, this can also be understood from the asymoptotics of the Mittag-Leffler function along the negative real axis \cite{paris2019asymptotics} and appendix~A. First, for $0<\alpha<1$, the algebraic long-time asymptotics (\ref{eq:A3b}) depends on $\alpha$ only via the amplitude which
is also seen in figure~\ref{fig:MittagLeffler_apetit}. On the other hand, for $1<\alpha\leq 2$
we rather need the more detailed expression (\ref{eq:A3c}) which for $\beta=1$ leads to 
\BEA
    E_{\alpha,1}\left(-D \frac{n^2 \pi^2}{L^2} t^\alpha\right) &\simeq& \frac{2}{\alpha} 
    \exp\left(\left(D \frac{n^2 \pi^2}{L^2}\right)^{{1}/{\alpha}} t \cos{\frac{\pi}{\alpha}}\right) \cos{\left( \left(D \frac{n^2 \pi^2}{L^2}\right)^{{1}/{\alpha}} t \sin{\frac{\pi}{\alpha}} \right)} 
    \nonumber \\
    & & - \sum_{m=1}^{\infty} \frac{1}{\Gamma(1-\alpha m)} 
    \left( -D \frac{n^2 \pi^2}{L^2}\right)^{-m} t^{-\alpha m}
    \label{26}
\EEA
where the trigonometric functions reflect the oscillatory behaviour seen in figure~\ref{fig:MittagLeffler_agrand}. 
For intermediate times, the last sum in (\ref{26}) is not yet the dominant term. Then the transient behaviour, as seen in figures~\ref{fig:shift2} and notably in figures~\ref{fig:shift4} and~\ref{fig:shift6} is determined by the terms in the first line of (\ref{26}), as follows: 

\subsubsection*{\underline{(a) Transient behaviour :}}

The term $\cos{\bigl( (D \frac{n^2 \pi^2}{L^2})^\frac{1}{\alpha} t \sin{\frac{\pi}{\alpha}} \bigr)}$ dominates, thus explaining the oscillatory behaviour of the temperature profiles  displayed in Figure 4 and 6. 

Besides, an oscillation period $T_n$ can be identified 
\begin{equation}
    \cos{\bigl( (D \frac{n^2 \pi^2}{L^2})^\frac{1}{\alpha} t \sin{\frac{\pi}{\alpha}} \bigr)} =:  \cos\left(\frac{2\pi t}{T_n}\right)
\end{equation}
which leads to 
\begin{equation}
    T_n = \frac{2\pi}{\sin \pi/\alpha} \left(D \frac{n^2 \pi^2}{L^2}\right)^{-\frac{1}{\alpha}} 
\end{equation}

\subsubsection*{\underline{(b) Asymptotic behaviour:} }

At longer times, the exponential term dominates and defines a time-scale $\tau$ for the damping
\begin{equation}
    \exp((D \frac{n^2 \pi^2}{L^2})^\frac{1}{\alpha} t \cos{\frac{\pi}{\alpha}}) =:  
    \exp\left(-\frac{t}{\tau}\right)
\end{equation}
such that explicitly 
\begin{equation}
    \tau = -\frac{1}{\cos \pi/\alpha}
    \left(D \frac{n^2 \pi^2}{L^2}\right)^{-\frac{1}{\alpha}} \geq 0
\end{equation}
For both times scales $T_n$ and $\tau$, the dependence on the physical parameters $D$ and $L$ is the same. 
Summarising this section, we have seen that in the fractional heat equation, while for $0<\alpha\leq 1$ the behaviour of the solutions is qualitatively analogous to simple diffusion, for $1<\alpha<2$ there is a passage from an oscillatory and wave-like behaviour for short times
to a damped diffusive behaviour at long times. 

\newpage
\section{The linear Pennes bioheat equation}\label{sec4}

Thermal ablation has emerged as a promising adjunctive therapy in the realm of cancer treatment \cite{baronzio2006hyperthermia}. This non-invasive approach consists in burning tumor cells by elevating the tissue temperature, but without damaging the surrounding healthy cells. One of the key elements underpinning the application of hyperthermia protoc\^ole in oncology is the Pennes Bioheat Equation, which consists in a local energy balance within a living tissue treated as a continuum with convective terms (blood perfusion) and internal sources (metabolic heat generation). This equation was originally proposed by C.K. Pennes in 1948 \cite{pennes_analysis_1948} and it can be expressed as:
\begin{equation} \label{1}
\rho c_t \frac{\partial T}{\partial t} = \nabla \cdot (k \nabla T) + Q_{\text{meta}} - \rho_b \omega_b c_b (T - T_{\text{b}})
\end{equation}
where $\rho$ is tissue density, $c_t$ is tissue specific heat, $T$ is temperature in the tissue,  $t$ is time,  $k$ is the associated thermal conductivity, $Q_{\text{meta}}$ metabolic heat generation, $\rho_b$ the blood density, $c_b$ the specific heat of the blood, $\omega_b$ is the blood perfusion rate and finally $T_{\text{b}}$ the arterial blood temperature.
This model has extensively been used to describe heat propagation in various biological tissue types such as liver \cite{peng2011two} and skin tissue \cite{yang_space-fractional_2021}. Ever since, refined versions of (\ref{1}) have been developed, taking into account more accurate descriptions of the blood flow (its direction for instance\cite{klinger1974heat} or space heterogeneity\cite{singh2024modified}). Improvements can also be introduced in the effective thermal conductivity such as in \cite{shih_effect_2002} or by considering detailed micro-vascular geometry \cite{weinbaum1984theory}. While this is certainly not an exhaustive list, it shows the strength of the Pennes model as a foundation upon which one can build an accurate model tailored for specific use cases. The call for more advanced models is all the more crucial with the study of thermal ablation applied to cancer therapy. Indeed, it was shown that tumor cells can develop thermo-tolerance if the thermal ablation is incomplete \cite{tomasovic1983heat}. As such, one can look towards including memory effects into the model so as to increase its accuracy and a prime candidate to do so would be fractional versions of the Pennes model. In 2013, Damor et al. \cite{damor_numerical_2013} replaced the original time derivative of the Pennes diffusion equation and numerically solved it for skin tissue. Ferr\'as {\it et al.} studied a dimensionally consistent version of the fractional Pennes model considered by Damor, in 2010 with a space dependant thermal diffusivity coefficient. 

In that same spirit, we replace the time-derivative $\partial_t$ by a fractional Caputo operator $\partial_t^{\alpha}$ and consider the following bioheat equation. 
\begin{equation} \label{pennes_frac}
\rho_t c_t \tau^{\alpha-1} \frac{\partial^{\alpha} T}{\partial t^{\alpha}} = \nabla \cdot (k \nabla T) + Q_{\text{meta}} - \rho c_b (T - T_{\text{b}})
\end{equation}
As displayed in this equation, a parameter $\tau [s]$ was added in order to solve the dimension inconsistency. From now on, 
we shall be working with a more compact form of the one written above, featuring reduced parameters:
\begin{equation}
\partial_t T = D \partial_x^2 T - D \gamma T  + \delta  \label{2}
\end{equation}
where
\begin{equation}
D = \frac{k}{\rho_t c_t \tau^{\alpha-1}} \;\; , \;\; D \gamma = \frac{\rho_b c_b \omega_b}{\rho_t c_t \tau^{\alpha-1}} \;\; , \;\;
\delta = \frac{Q_{\rm meta}}{\rho_t c_t \tau^{\alpha-1}} + \frac{\rho_b c_b \omega_b}{\rho_t c_t \tau^{\alpha-1}} T_b
\end{equation}

This inhomogeneous equation can be reduced to the homogeneous case $\delta=0$ by substituting
$T \mapsto T - \delta/D\gamma$. In the following, we then solve the homogeneous Pennes equation by means of the separation ansatz
$T(t,x) = f(t) g(x)$. This leads to the two independent equations
\begin{equation}
\partial_t^{\alpha} f(t) = -D k^2 f(t) \;\; , \;\;
g''(x) = -\bigl( k^2 - \gamma\bigr) g(x)
\end{equation}
where $k$ is the separation constant. The space-dependent part now becomes
$g(x) = g_0 e^{\sqrt{\gamma - k^2\,}\, x} + g_1 e^{-\sqrt{\gamma - k^2\,}\, x}$ whereas the time-dependent
part can be taken over from our previous treatment of eq.~(\ref{time}). This gives the formal solution (for $0<\alpha<2$) 
\begin{eqnarray} \label{eq:Tansatz5} 
T(t,x) &=& \sum_k \left[ f_0(k) E_{\alpha,1}(-D k^2 t^{\alpha}) e^{\sqrt{\gamma - k^2\,}\, x} 
+  f_1(k) E_{\alpha,1}(-D k^2 t^{\alpha}) e^{-\sqrt{\gamma - k^2\,}\, x} \right. \nonumber \\
& & \left. + f_0'(k)  tE_{\alpha,2}(-D k^2 t^{\alpha}) e^{\sqrt{\gamma - k^2\,}\, x} 
+  f_1'(k) tE_{\alpha,2}(-D k^2 t^{\alpha}) e^{-\sqrt{\gamma - k^2\,}\, x} \right] 
\nonumber \\
& & 
\end{eqnarray} 
We shall consider the situation where we heat the middle of our domain (of total length $L=20cm$) at a fixed temperature $T_h$.
Thus, will be using the following initial/boundary conditions for this part
\begin{equation}  \label{inibord_pennes}
    \begin{cases}
    T(t,0)=0\\
    T(t,L)=0 \\
    T(0,x)=\frac{T_h}{L^2}\left(x-L/2\right)^2+T_h
    \end{cases}
\end{equation}
and also assume a stationary initial distribution. 
The first of these implies the conditions
\begin{equation}
f_0(k) + f_1(k) =0 \;\; , \;\; f_0'(k) + f_1'(k) =0
\end{equation}
Inserting this into (\ref{eq:Tansatz5}), the second boundary condition leads to, for all $x\in[0,L]$ 
\begin{equation}
e^{\sqrt{\gamma - k^2\,}\, x} - e^{-\sqrt{\gamma - k^2\,}\, x} \stackrel{!}{=} 0
\end{equation}
which can also be written as $\sin \bigl(\sqrt{k^2-\gamma\,}\,L\bigr)\stackrel{!}{=}0$.  Hence we have the admissible values
\begin{equation}
k^2 = k_m^2 = \left( \frac{m\pi}{L}\right)^2 + \gamma \;\; ; \;\; m\in\mathbb{N}
\end{equation}
such that 
\BEQ \label{eq:Tansatz6} 
T(t,x)=  
\sum_{m=1}^{\infty} \left[ f_m E_{\alpha,1}(-D \left( \frac{m^2 \pi^2}{L^2} +\gamma\right) t^{\alpha}) 
+ f_m'  tE_{\alpha,2}(-D \left( \frac{m^2 \pi^2}{L^2} +\gamma\right) t^{\alpha})  \right]
\sin\left( m\pi \frac{x}{L}\right) \nonumber \\
\EEQ 
The initial conditions now fix the values of $f_m$ and $f_m'$ as shown in ~\ref{sec2}.  
For convenience sake, the Fourier coefficients were found for an initial profile defined on the domain $[-L,L]$. 
Doing so will fix $f_m = \frac{4 T_h}{(m \pi)^2} (-1)^{m-1}$ and $f_m'=0$.

The final solution of   (\ref{pennes_frac}) with the initial/boundary conditions (\ref{inibord_pennes}) is, 
after restituting the in-homogeneous constant term (valid for $0<\alpha<2$ for a stationary initial condition) 
\begin{eqnarray} \label{eq:Tansatz7} 
T(t,x) &=& T_b + \frac{2 T_h}{3}+\frac{Q_meta}{\rho_b c_b \omega_b } \\
& & + 
\sum_{m=1}^{\infty}  \frac{4 T_h}{(m \pi)^2} (-1)^{m-1}  E_{\alpha,1}\left(-D \left( \frac{m^2 \pi^2}{L^2} +\gamma\right) t^{\alpha}\right)  
\cos\left( \frac{m\pi}{L} \left(2 x+L\right)\right) \nonumber 
\end{eqnarray} 
We can then showcase the solution to the fractional Pennes bioheat equation. The values of the parameters such as $T_b, \rho_b, c_b$ and such, 
will be taken  from various papers investigating bioheat models tailored after several biological tissue types (for instance in the prostate \cite{kabiri2019theoretical}, 
skin \cite{yang2021space} or liver\cite{barnoon2020magnetic}).

\begin{figure}[H]
    \centerfloat
    \includegraphics[scale=0.35]{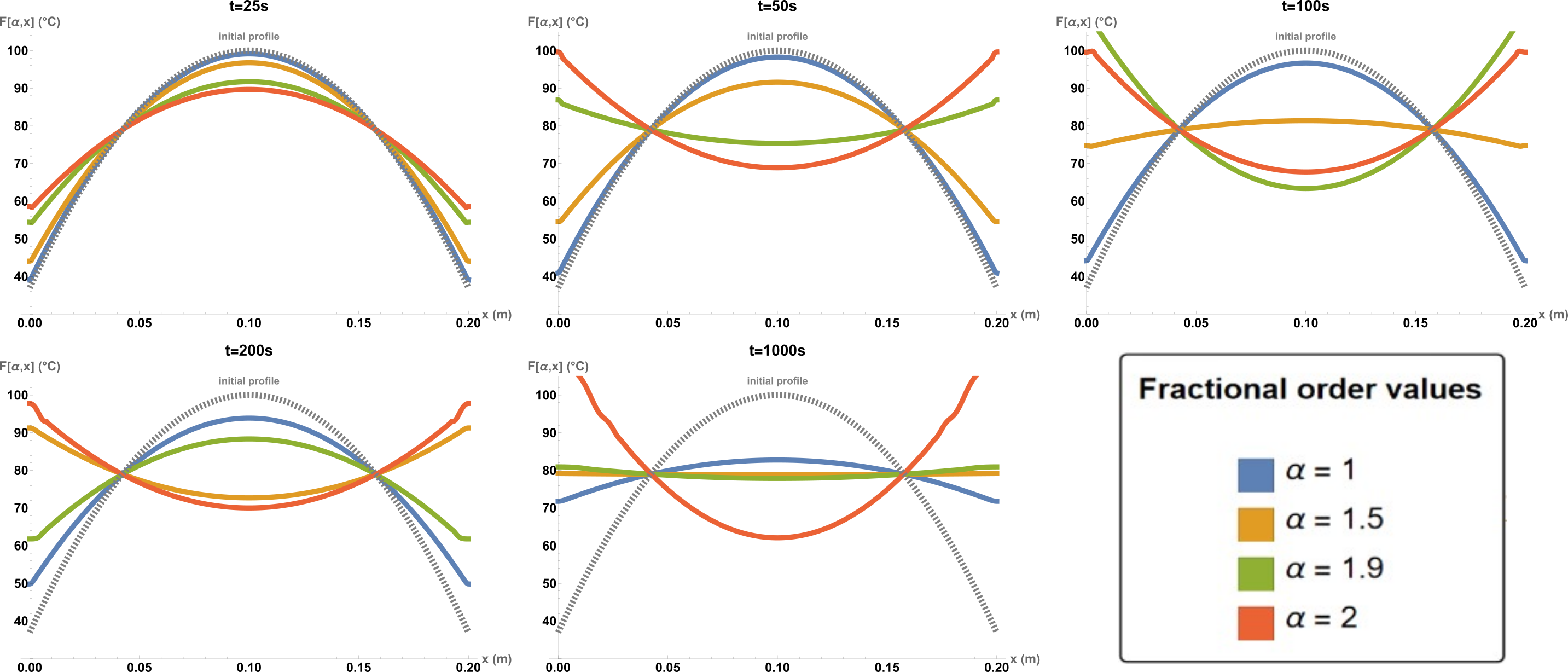}
    \caption{Temperature profile $T(t,x)$ at $t=[25s, 50s, 100s, 200s, 1000s]$ for a quadratic initial profile. 
    We consider here, a temperature increase to $100$°C at the centre of our domain of length $20cm$. 
    The plots were generated for $500$ terms in the Fejer summation for values of $\alpha=[1,1.5,1.9,2]$.
    }
    \label{fig:pennes_x2_high}
\end{figure}

\begin{figure}[H]
    \centerfloat
    \includegraphics[scale=0.25]{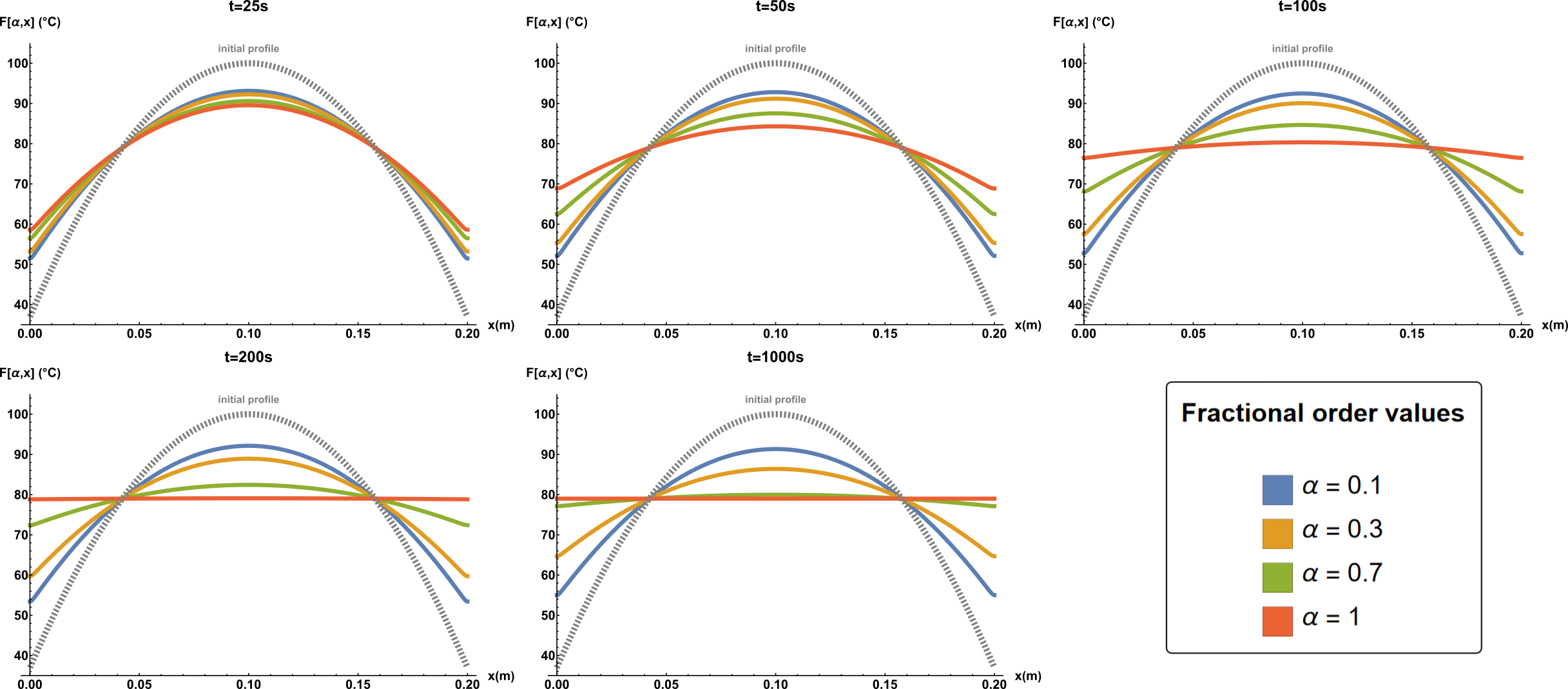}
    \caption{Temperature profile $T(t,x)$ at different simulation times $t$ for a quadratic initial profile. 
    We consider here, a temperature increase to $100$°C at the centre of our domain of length $10cm$. 
    The plots were generated for $500$ terms in the Fejer summation for values of $\alpha=[0.1,0.3,0.7,1]$.
    }
    \label{fig:pennes_x2_low}
\end{figure}

The observations that were formulated in section \ref{sec2} are also underlined here with this applied case, as the solutions display a transition between oscillatory regime at low values of $t$ to a diffusive regime for higher values of time $t$.
The addition of the Pennes terms here do not introduce significant changes compared to what had previously been highlighted in the previous section, for this specific set of initial/boundary conditions. The term $\frac{Q_m}{\rho_b c_b \omega_b }$ resulting from the convection aspect of the model, will simply offset the temperature profiles by a certain value. 

However, the influence of the terms specific to the Pennes model can be better observed when considering initial profiles such as the linear one from the previous section.

\begin{figure}[H]
    \centerfloat
    \includegraphics[scale=0.7]{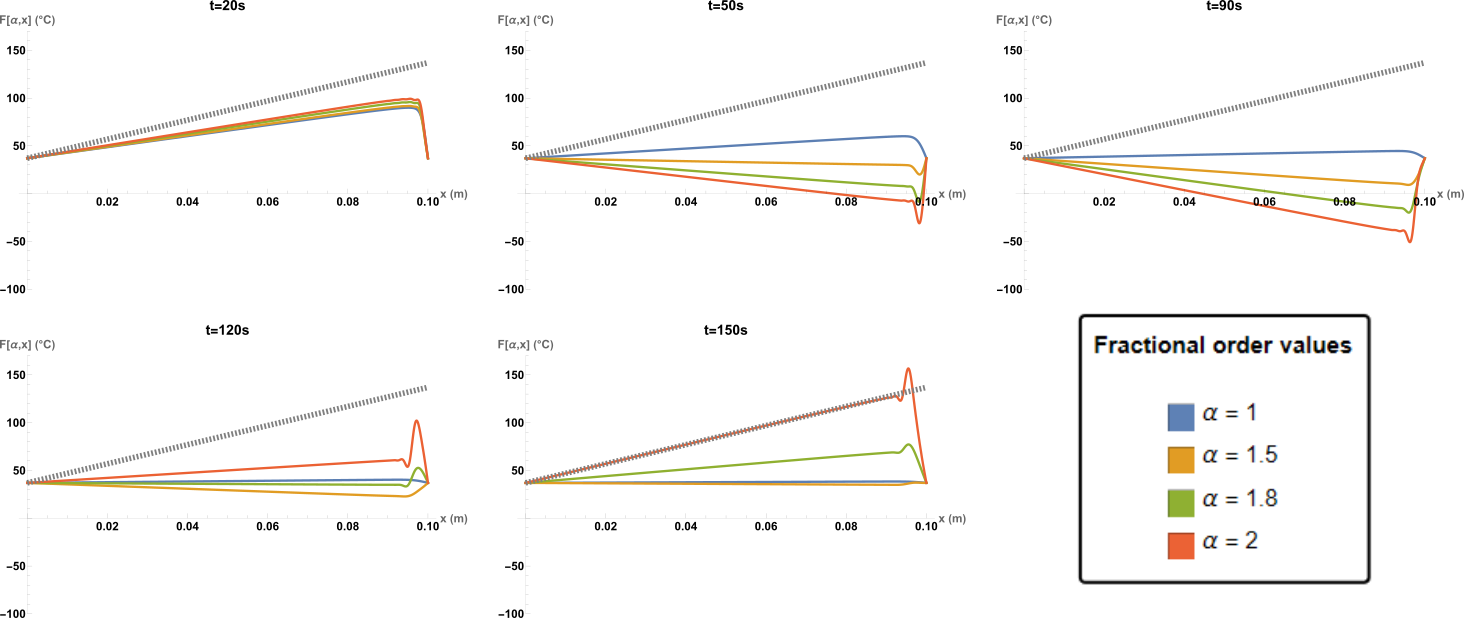}
    \caption{Temperature profile $T(t,x)$ at different simulation times $t$ for a linear initial profile (dashed gray). We consider here, a linear temperature increase to $100$°C our domain of length $10cm$. The plots were generated for $200$ terms in the Fejer summation for values of $\alpha$ ranging from $1$ to $2$
    }
    \label{fig:pennes_lin_high}
\end{figure}

As can be seen from these figures, the temperature profiles displayed here seem to oscillate back and forth as time increases, a behaviour which had previously not observed in \ref{sec2}. 

In fact, this swinging motion of the slopes can be understood, should one take a closer look at the solution to \eqref{pennes_frac}, specifically for $\alpha=2$.

\begin{figure}[H]
    \centerfloat
    \includegraphics[scale=0.5]{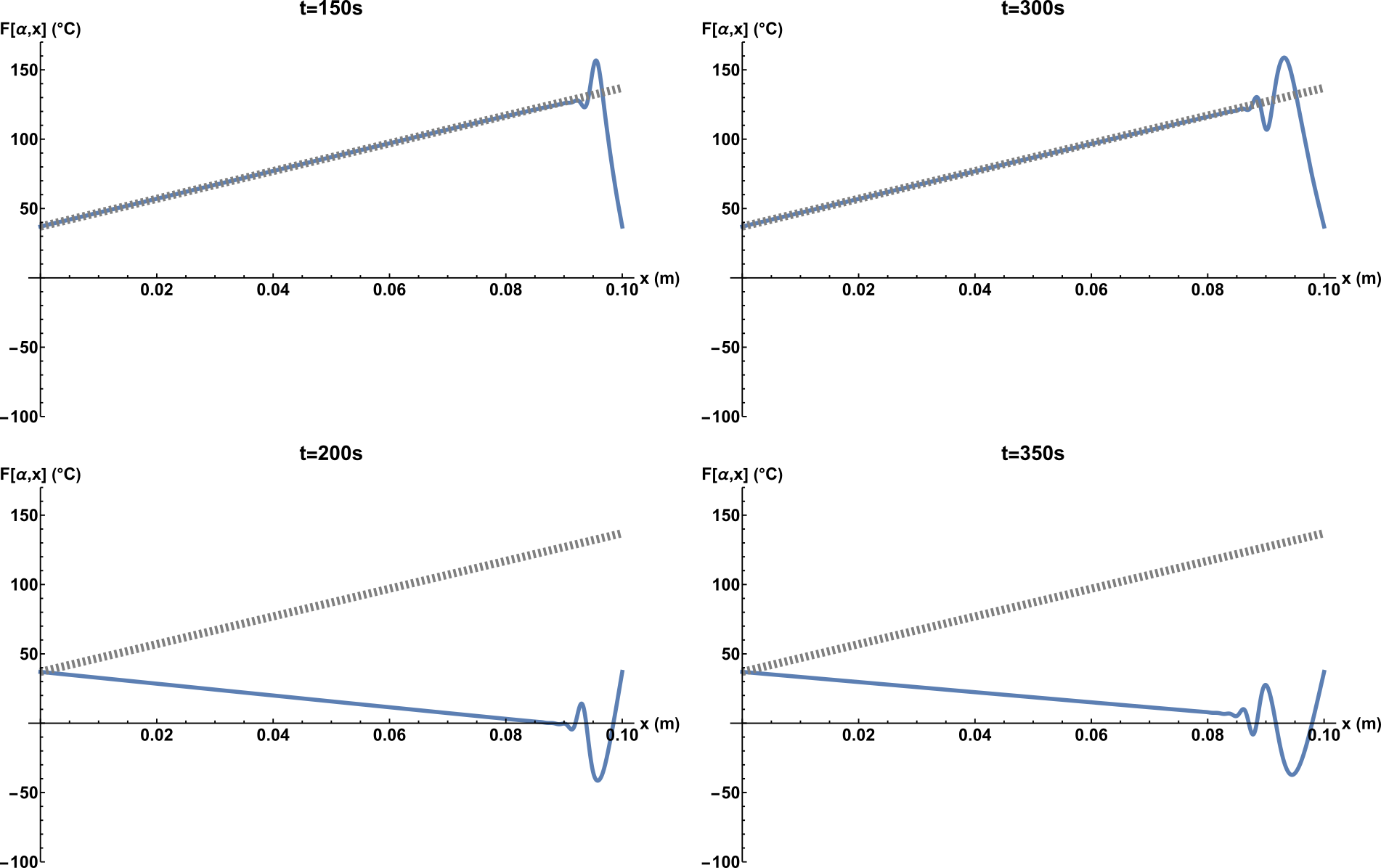}
    \caption{Temperature profile $T(t,x)$ at simulation times $t$ and $t+p_0$ for $\alpha=2$. 
    }
    \label{fig:pennes_period}
\end{figure}

It can be shown that for this value of the fractional parameter, the solution is :
\begin{equation} \label{pennes_alpha2}
T(t,x) = \frac{200}{\pi}\sum_{m=1}^{\infty}   \frac{(-1)^{m-1}}{m}  \cos\left(D^{\frac{1}{2}} \left( \frac{m^2 \pi^2}{L^2} +\gamma\right)^{\frac{1}{2}} t\right)  
\sin\left( \frac{m\pi}{L} x\right)
\end{equation}
If we take a look at the cosine term, when considering the case where $\gamma>>\frac{m^2 \pi^2}{L^2}$ 
(which is generally the case considering the value of $\gamma$. It is about $10^6$ using the values from \cite{barnoon2020magnetic})
we can derive the following results :
\begin{eqnarray} \label{pennes_alpha2_approx} 
\cos\left(D^{\frac{1}{2}} \left( \frac{m^2 \pi^2}{L^2} +\gamma\right)^{\frac{1}{2}} t\right) &=& \cos\left((D\gamma)^{\frac{1}{2}}\right) \cos\left(\frac{m^2 \pi^2}{L^2}t\right)\\ \nonumber
& & - \sin\left(\left(D\gamma\right)^{\frac{1}{2}}\right)\sin\left(\frac{m^2 \pi^2}{L^2}t\right)
\end{eqnarray} 
In the end 
\begin{equation}
    \cos\left(D^{\frac{1}{2}} \left( \frac{m^2 \pi^2}{L^2} +\gamma\right)^{\frac{1}{2}} t\right) \equiv \cos\left(\frac{2\pi}{p_0}t\right)\cos\left(\frac{2\pi}{p_m}t\right)-\sin\left(\frac{2\pi}{p_0}t\right)\sin\left(\frac{2\pi}{p_m}t\right)
\end{equation}
where we can derive certain periods : 
\begin{equation}
    \begin{cases}
        p_0=2\pi\dfrac{\rho_t c_t\tau}{\rho_b c_b \omega_b}\\
        \\
        p_m = \dfrac{4L^2\gamma^{\frac{1}{2}}}{m^2\pi}, \forall n\in \mathbb{N}^*
    \end{cases}
\end{equation}
Using parameter values corresponding to the investigation on liver cancer\cite{barnoon2020magnetic} with $\tau=16s$ \cite{kabiri2019theoretical}, 
we derive $p_o=152s$ which in line with the time it takes for the temperature profile to swing back and forth, as highlighted by Figure\ref{fig:pennes_period}

\section{Concluding remarks} 

In this paper, we present a mean to derive the exact analytical solution to a fractional diffusion equation using the separation of variable method, 
as well as the properties of the Mittag-Leffler function. The solutions were calculated for values of the fractional order $\alpha$ 
ranging from 0 to 2, for various sets of boundary and initial conditions including a linear, quadratic and quartic initial profiles. 
The results highlight a transition between an oscillatory regime at low time $t$, to a diffusive regime for larger values of $t$. Furthermore, 
aforesaid properties were better emphasised for initial profiles displaying rapid weight decays with respect to their Fourier decompositions, 
such as the quadratic and quartic profiles. Eventually, we conclude with an application of our method to a fractional Pennes bioheat model. 
Not only are the previous observations from section~\ref{sec3} still valid in this last section~\ref{sec4}, 
we also point out the appearances of global slope swings resulting from the convective terms exclusive to the Pennes model. 

These preliminary results call for further developments of the present work. 
These latter include the study of diffusion processes involving both fractional time and space derivatives. Although formal and general solutions have already been proposed by Mainardi \cite{gorenflo2000wright,mainardi2001fundamental}, 
the physical content of these solutions is yet to be disclosed. In particular, a challenging pathway to explore would be to investigate the physics of a fractional Maxwell-Cattaneo heat conduction model. Indeed, the regular diffusion equation is known to suffer from severe flaws, such as an infinite-speed heat conduction. In the framework of Fourier heat conduction, this issue is fixed by adding a first order time derivative to the phenomenological Fourier law \cite{christov2005heat}. But the way to implement this correction in the case of fractional heat diffusion does not rally consensus: following Compte and Metzler \cite{compte1997generalized}, there are indeed at least four ways of extending Cattaneo works, depending on physical assumptions underlying the model (continuous time random walks, delayed flux-force relation, non-local transport theory...). The outcomes for biophysics and thermal ablation in particular still remain to be clarified.

\vspace{5cm}
\noindent {\bf{Acknowledgements:}}
This work was supported by  the  Coll\`ege Doctoral %
``Statistical Physics of Complex Systems'' Leipzig-Lorraine-Lviv-Coventry (${\mathbb L}^4$). 

\newpage
\appsection{A}{The Mittag-Leffler function}

We provide some background on the Mittag-Leffler function  \cite{Mainardi1996,podlubny_fractional_1998,Diethelm10,Haubold11,Gorenflo14,Giusti22,vanMieghem2023}.  It is an entire function, defined  for $z\in\mathbb{C}$ by 
\begin{equation}
E_{\alpha,\beta}(z) = \sum_{n=0}^{\infty} \frac{z^n}{\Gamma(\alpha n +\beta)}
\end{equation}
where $\alpha,\beta$ are real and positive constants and $\Gamma$ is Euler's Gamma function. From the definition, one directly has
\BEQ
E_{1,1}(-x^2) = e^{-x^2} \;\; , \;\; E_{0,1}(-x^2) = \frac{1}{1+x^2}
\EEQ
such that $E_{\alpha,1}(-x^2)$ interpolates for $0<\alpha<1$ between a gaussian and a lorentzian distribution. 
One has 
\begin{eqnarray}
E_{\alpha,\beta}(0)&=& \frac{1}{\Gamma(\beta)} \nonumber \\
E_{\alpha,\beta}(z)&=&\beta E_{\alpha,\beta+1}(z) + a z \frac{\D}{\D z} E_{\alpha,\beta+1}(z) \label{ML-recur}  \\
\frac{\D^m}{\D z^m} \left( z^{\beta-1} E_{\alpha,\beta}(z^{\alpha}) \right)&=&z^{\beta-m-1} E_{\alpha,\beta-m}(z^{\alpha}) \;\; ; \;\;
\beta-m >0   \nonumber 
\end{eqnarray}
Along the real axis, the leading asymptotic expansions read, for $0<\alpha<2$ \cite[eqs. (6.10,6.11)]{Haubold11}, \cite[eq. (4.25)]{Giusti22} 
\begin{figure}[H]
    \centerfloat
    \includegraphics[scale=0.6]{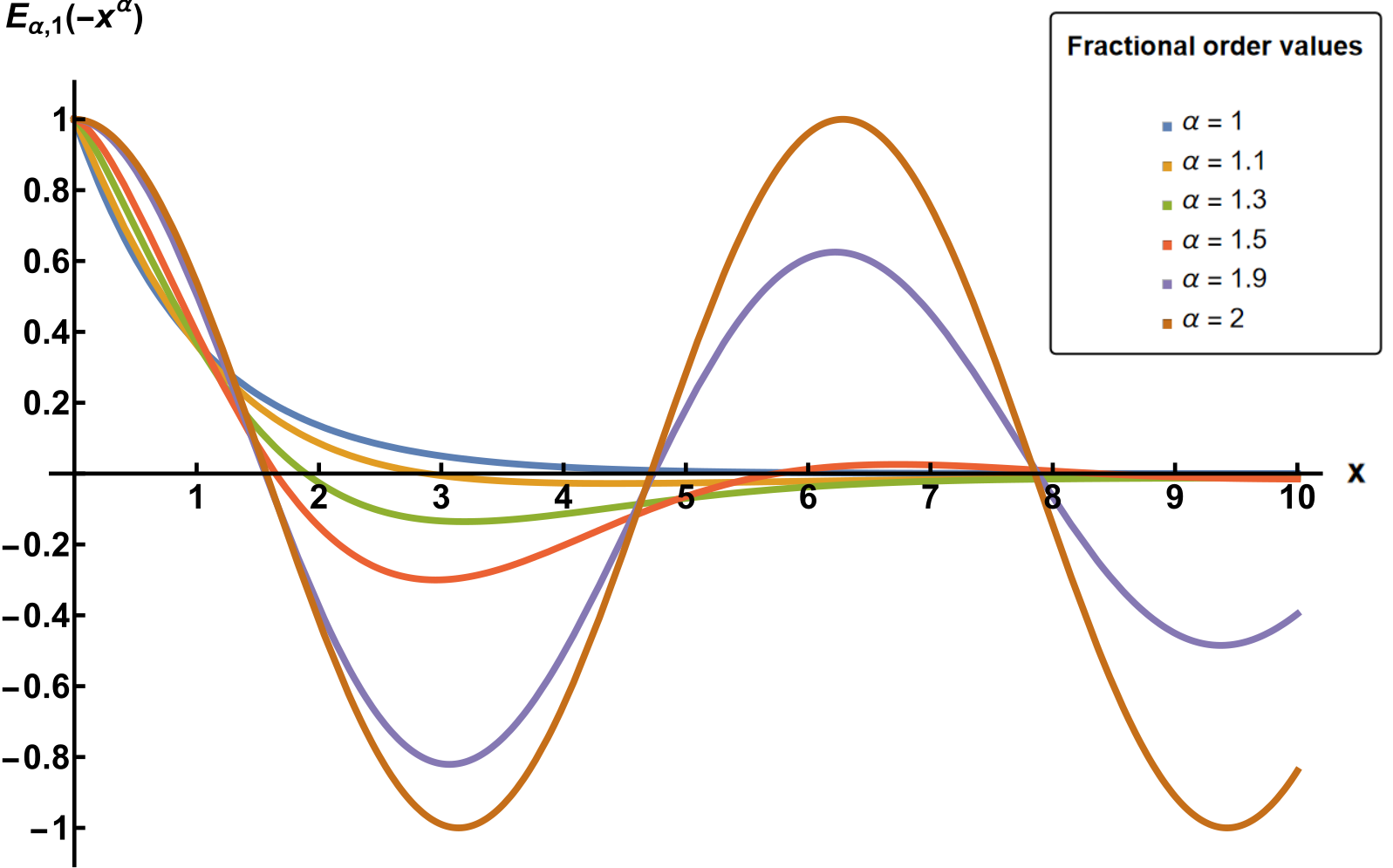}
    \caption[MLpetit]{Mittag-Leffler function $\mathscr{E}_{\alpha}(x) := E_{\alpha,1}(-x^{\alpha})$ for $\alpha=[1,1.1,1.3,1.5,1.9,2]$.
    }
    \label{fig:MittagLeffler_agrand}
\end{figure}
\begin{subequations} \label{eq:A:Easy}
\begin{align}
E_{\alpha,\beta}(x) &\simeq \frac{1}{\alpha} x^{(1-\beta)/\alpha} e^{x^{1/\alpha}} \left( 1 +  {\rm O}(x^{-1})\right)  \label{eq:A3a} \\
E_{\alpha,\beta}(-x) &\simeq \frac{1}{\Gamma(\beta-\alpha)} \frac{1}{x} \left( 1 +  {\rm O}(x^{-1})\right) 
\;\; ; \;\; \beta\ne \alpha  \label{eq:A3b} 
\end{align} 
Along the negative real axis, the asymptotic behaviour for $x\to\infty$ can be stated more precisely for $1<\alpha<2$ as \cite{paris2019asymptotics}
\begin{align}
E_{\alpha,\beta}(-x) &\simeq \frac{2}{\alpha} x^{({1-\beta})/{\alpha}}  
    \exp\left(x^{{1}{/\alpha}} \cos{\frac{\pi}{\alpha}}\right) \cos\left(x^{{1}/{\alpha}}\sin{\frac{\pi}{\alpha}}+\frac{\pi(1-\beta)}{\alpha}\right) 
    \nonumber \\
    & ~~- \sum_{n=1}^{\infty} \frac{(-x)^{-n}}{\Gamma(\beta-\alpha n)}  \label{eq:A3c}
\end{align}
\end{subequations}
and shows that at least for $\alpha>1$, there is a regime at not too large values of $|x|$ where the function $E_{\alpha,\beta}(-x)$ oscillates as a function of $x$. This is shown in figure~\ref{fig:MittagLeffler_agrand} for the range $1\leq \alpha\leq 2$. {\it A contrario}, the Mittag-Leffler function is positive and strictly decreasing for $0<\alpha\leq 1$, as illustrated in figure~\ref{fig:MittagLeffler_apetit}. 
\begin{figure}[tb]
    \centerfloat
    \includegraphics[scale=0.6]{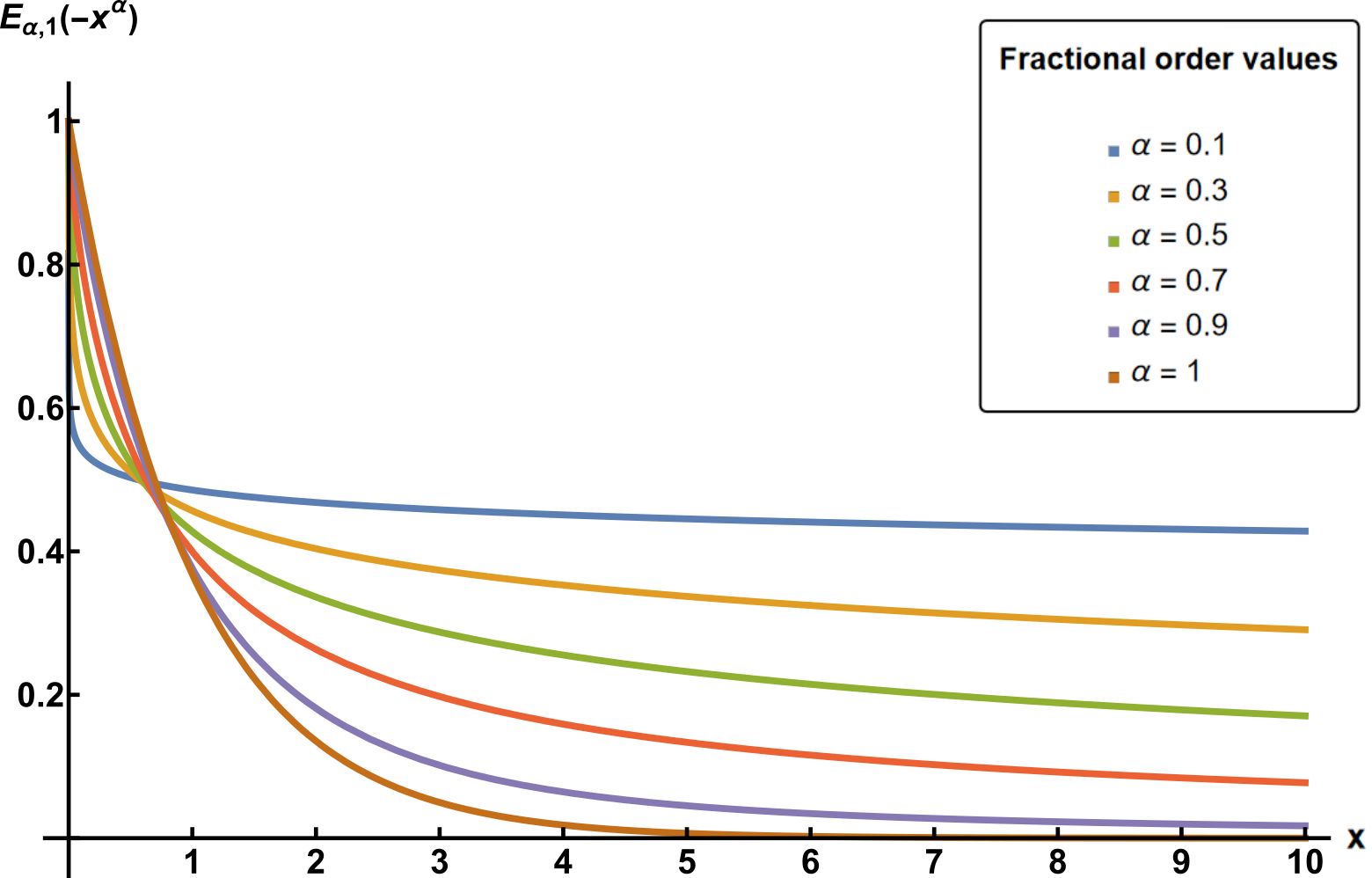}
    \caption[MLgrand]{Mittag-Leffler function $\mathscr{E}_{\alpha}(x) := E_{\alpha,1}(-x^{\alpha})$ for $\alpha=[0.1,0.3,0.5,0.7,0.91]$.
    }
    \label{fig:MittagLeffler_apetit}
\end{figure}

In addition, one has the Laplace transform \cite{Podlubny1997,podlubny_fractional_1998}
\begin{equation} \label{eq:lapML}
\mathscr{L}\left( t^{\alpha n +\beta -1} \frac{\D^n}{\D t^n} E_{\alpha,\beta}\bigl(\pm c t^{\alpha}\bigr)\right)(p) 
= \int_0^{\infty} \!\D t\: e^{-pt} t^{\alpha n +\beta -1} \frac{\D^n}{\D t^n} E_{\alpha,\beta}\bigl(\pm c t^{\alpha}\bigr) 
= \frac{n!\: p^{\alpha-\beta}}{\bigl( p^{\alpha}\mp c\bigr)^{n+1}}
\end{equation}
Some information is available on the zeroes of the function $E_{\alpha}(x) := E_{\alpha,1}(-x^{\alpha})$ \cite{Diethelm10}, which goes beyond the asymptotic form (\ref{eq:A3c}). 
For $0<\alpha\leq 1$, $E_{\alpha}(x)>0$ for all $x>0$. For $1<\alpha<2$ one has
the following: 1. if $\alpha=1+\vep$, the zero with the smallest absolute value of $E_{\alpha}(x)$ 
occurs at $x^*\simeq \ln\frac{2}{\vep}$. 2. if $\alpha=2-\vep$, the zero with the largest
absolute values occurs at $x^*\simeq \frac{12}{\pi \vep \ln \vep}$.

\newpage 
\appsection{B}{Techniques for solving partial differential equations}
We briefly recall several standard techniques for solving {\em partial} differential equations, 
in semi-infinite space, which might
become useful when studying fractional derivatives. Our paradigmatic example is the heat equation
\BEQ \label{eq:B1}
\partial_t u = D \partial_x^2 u \;\; , \;\; u(t,0)=1 \;\; , \;\; \partial_x u(t,0) = 0 \mbox{\rm ~~and~~} u(0,x)=T
\EEQ
for all times $t\geq 0$ and in the entire half-space $x\geq 0$. 

\subsection{Scaling ansatz} 

For equations with a natural scale-invariance, one might seek for solutions of the form
\BEQ \label{eq:scal}
u(t,x) = t^{\beta} f\bigl( x t^{-1/2}\bigr)
\EEQ
which should work for large variables, $t\to\infty$ and $x\to\infty$ such that the scaling variable
$\xi := x t^{-1/2}$ is kept fixed. Compatibility  of the scaling ansatz (\ref{eq:scal}) with the differential equation implies (\ref{eq:B1}) that $\beta=0$ and the two limits
\BEQ \label{gl:lim}
f(0) = 1 \;\; , \;\; f(\infty) = T
\EEQ
The attractive feature is that eq.~(\ref{eq:B1}) reduces to an
ordinary differential equation, namely
\BEQ \label{eq:f}
D f''(\xi) + \frac{1}{2} \xi f'(\xi) = 0
\EEQ
for the scaling function $f$. The solution of (\ref{eq:f}) is readily found to be 
\BEQ
f(\xi) = f_0 + g_0 \int_0^{\xi} \!\D\xi'\: e^{-\xi'^2/4D} = f_0 + g_0 \sqrt{\pi D\,}\, \erf\left( \frac{\xi}{\sqrt{4D\,}}\right)
\EEQ
where $\erf$ is the error function and the prescribed limit behaviour (\ref{gl:lim}) 
fixes the constants $f_0$, $g_0$. One finally has
\BEQ \label{gl:final}
u(t,x) = f(\xi) = 1 + \bigl( T-1\bigr) \erf \left(\frac{\xi}{2\sqrt{D}}\right) 
       = 1 + \bigl( T-1\bigr) \erf\left( \frac{x}{\sqrt{4D\, t\,}} \right)
\EEQ
This satisfies two of the three boundary conditions in (\ref{eq:B1}). However, since $f'(0)=\frac{T-1}{\sqrt{\pi D\,}}$, the last  boundary condition $\partial_x u(t,0)=0$ is only satisfied in the special case $T=1$. 

\subsection{Domain extension}
This is quite analogous to the method of images in electrostatics. One first formally extends \cite{Durang10}
eq.~(\ref{eq:B1}) from the half-line formally to the entire line $x\in\mathbb{R}$ by defining
\BEQ \label{gl:defneg}
u(t,-x) := 2 - u(t,x) \mbox{\rm ~~~for $x>0$}
\EEQ
After the extension, one considers the Fourier transformation
\BEQ
\wht{u}(t,k) = \frac{1}{\sqrt{2\pi\,}} \int_{\mathbb{R}} \!\D x\: e^{-\II k x} u(t,x)
\EEQ 
and admits the boundary conditions $u(t,\pm\infty) = \partial_x u(t,\pm\infty)=0$. The  Fourier-transformed
equation reads $\partial_t \wht{u} = - D k^2 \wht{u}$. It has the solution
\BEQ
\wht{u}(t,k) = \wht{u}(0,k) e^{-D k^2 t}
\EEQ
The Fourier back-transformation is standard and leads to
\BEQ
u(t,x) = \frac{1}{\sqrt{4\pi D\,t\,}} \int_{\mathbb{R}} \!\D y\: u(0,y) e^{-(x-y)^2/4Dt}
\EEQ
For $y<0$, one now uses the definition (\ref{gl:defneg}) such that
\BEA
u(t,x) &=& \frac{1}{\sqrt{4\pi D\,t\,}} \int_0^{\infty} \!\D y\:
\left( u(0,y) e^{-(x-y)^2/4Dt} + u(0,-y) e^{-(x+y)^2/4Dt} \right) 
\nonumber \\
&=& \frac{2}{\sqrt{4\pi D\,t\,}} \int_0^{\infty} \!\D y\: e^{-(x+y)^2/4Dt} \\
& & + \frac{1}{\sqrt{4\pi D\,t\,}} \int_0^{\infty} \!\D y\:  u(0,y) \left( e^{-(x-y)^2/4Dt} - e^{-(x+y)^2/4Dt} \right) \nonumber 
\EEA
which gives the general solution for any prescribed initial value $u(0,x)$. 
Now, imposing the initial condition $u(0,y)=T$ for all $y>0$ leads back to the end solution (\ref{gl:final}). 

\subsection{Laplace transform}

For the application of the Laplace transformation method, one needs the following. 

\noindent
{\bf Lemma.} {\it If the function $f:\mathbb{R}_+ \to \mathbb{R}_+$ is absolutely integrable, then for any $p\geq 0$ the Laplace transform
\BEQ 
\bigl( \mathscr{L} f\bigr) (p) = \lap{f}(p) = \int_0^{\infty} \!\D t\: e^{-p t} f(t)
\EEQ
is a non-increasing function of $p$.} 

\noindent
{\bf Proof:} Let $p'>p>0$. Then $\lap{f}(p') = \int_0^{\infty} \!\D t\: e^{-p' t} f(t) \leq \int_0^{\infty} \!\D t\: e^{-p t} f(t) = \lap{f}(p)$. \hfill q.e.d. \\

For $x>0$, the Laplace-transformed equation (\ref{eq:B1}) becomes $p \lap{u}(p,x) -T = D \partial_x^2\lap{u}(p,x)$. Since a special solution is $\lap{u}_{\rm sp}=\frac{T}{p}$, the general solution of this is
\BEQ 
\lap{u}(p,x) = u_{+} e^{+\sqrt{p/D\,}\, x} + u_{-} e^{-\sqrt{p/D\,}\, x} + \frac{T}{p}
\EEQ 
with free constants $u_{\pm}$. 
Herein, the first term is an increasing function of $p$ and cannot be interpreted as a Laplace-transform. We discard it by setting $u_+=0$. The boundary condition $u(t,0)=1$ leads to $\lap{u}(p,0)=1/p$, hence
$u_{-}=(1-T)/p$ and
\BEQ
\lap{u}(p,x) = \frac{1-T}{p} e^{-\sqrt{p/D\,}\, x} + \frac{T}{p}
\EEQ
The inverse Laplace-transform of this, using \cite[(2.2.1.16)]{Prudnikov5}, leads back to (\ref{gl:final}). 

\section*{References}
\providecommand{\newblock}{}

\end{document}